\documentclass[final,1p,times]{elsarticle}

\usepackage{amssymb}
\usepackage{lipsum}

\usepackage{epsfig,color}
\setcitestyle{square}
\usepackage{amsmath}

\journal{Physica D}

\begin{document}

\begin{frontmatter}



\title{
Lagrangian Homotopy Analysis Method 
using the Least Action Principle 
}


\author[first]{Gervais Nazaire Chendjou Beukam}
\author[first]{Jean Pierre Nguenang} 
\author[second,third,fourth]{Stefano Ruffo}
\author[second,third,sixth]{Andrea Trombettoni}

\affiliation[first]{organization={Pure Physics Laboratory: Group of Nonlinear Physics and Complex Systems, Department of Physics},
            addressline={Faculty of Science, University of Douala}, 
            city={Douala},
            postcode={Box 24157}, 
            country={Cameroon}}

\affiliation[second]{organization={SISSA},
            addressline={Via Bonomea 265}, 
            city={Trieste},
            postcode={I-34136}, 
            country={Italy}}
            
\affiliation[third]{organization={INFN},
            addressline={Sezione di Trieste}, 
            city={Trieste},
            postcode={I-34151}, 
            country={Italy}}
            
\affiliation[fourth]{organization={Istituto dei Sistemi Complessi, Consiglio Nazionale delle Ricerche},
            addressline={via Madonna del Piano 10}, 
            city={Sesto Fiorentino},
            postcode={I-50019}, 
            country={Italy}}
\affiliation[sixth]{organization={Department of Physics, University of Trieste},
            addressline={Strada Costiera 11}, 
            city={Trieste},
            postcode={I-34151}, 
            country={Italy}}

\begin{abstract}
The Homotopy Analysis Method (HAM) is a powerful 
technique which allows to derive approximate solutions of both ordinary and partial differential equations. We propose to use a variational approach based on the Least Action Principle (LAP) in order to improve the efficiency of the HAM when applied to 
 Lagrangian systems.
The extremization of the action is achieved by varying the HAM parameter, 
therefore controlling the accuracy of the approximation. As case studies we consider 
the harmonic oscillator, the cubic and the quartic anharmonic oscillators, 
and the Korteweg-de Vries partial differential equation. We compare our results with those obtained using the standard approach, which is based on the residual error square method. We 
see that our method accelerates the convergence of the HAM parameter 
to the exact value in the cases in which the exact solution is known. When the exact solution is not analytically known, we find that our method performs better than the standard HAM for the cases we have analyzed. Moreover, our method shows better performance when the order of the approximation is increased and when the nonlinearity of the equations is stronger.
\end{abstract}






\end{frontmatter}




\section{Introduction}
\label{introduction}

Highly nonlinear problems require sophisticated numerical methods and a vast variety of techniques has been developed to find 
approximate solutions of nonlinear ordinary differential equations (ODE) and partial differential equations (PDE). Among the various approaches, one characterized 
by a wide range of applicability is provided by the Homotopy Analysis Method (HAM)~\citep{SJLiao2003,SJLiao2014}. It can be typically used when other approaches fail to provide the 
desired results \citep{Sanz-serna1994}. The method employs homotopy, a 
concept in topology, to generate convergent series solutions of nonlinear systems. This semi-analytical method provides a viable alternative to 
other techniques such as the Lyapunov’s artificial small parameter method \cite{Lyapunov1992}, the Adomian decomposition method 
\cite{Adomian1991}, the Delta expansion method 
\cite{Ji-huan2002}, the homotopy perturbation method 
\cite{Ji-huan1999}, and 
in general to techniques that cannot guarantee the convergence of the series of the approximate solutions and are valid only for weakly nonlinear problems. 
In the HAM, 
convergent series solutions can be obtained and used 
also in presence of a strong 
nonlinear perturbation. 
The HAM maintains one fundamental aspect of perturbation theory, which is the fact that one may iteratively solve the equations, 
and 
at the same time it does not require a small parameter. 
Unlike other analytical approximation techniques, the HAM provides 
a flexible and convenient way to guarantee the convergence of the series which approximates the solution by means of introducing an auxiliary parameter, the so-called {\it HAM parameter}, usually denoted by $c_0$. This parameter, which must be non zero in order to treat nonlinear problems, is a 
variable which provides a simple way to enforce the convergence of the approximate solution. 
As a result, the HAM is generally valid for various types of equations with high-order nonlinearity, especially for those where a small parameter cannot be identified. 
In the frame of the HAM, one has 
freedom to choose the auxiliary linear operator, 
the initial guess of the solution 
and the value of $c_0$. 
It should be 
emphasized that it is 
not possible in general to introduce 
a control parameter like $c_0$ in the frame of perturbative techniques, like the Lyapunov's small parameter method \cite{Szezech2004}. When the auxiliary linear operator is properly chosen, the value of the HAM parameter $c_0$ appears to stabilize around a certain fixed value as the order of the approximation is increased. 

In 
\cite{Liao2010} it was proposed to use the residual error square technique in order to accurately determine the HAM parameter $c_0$. The residual error square (see below in 
section~\ref{method}) characterizes the global error between the 
approximation obtained by HAM and the exact solution. 
We introduce here a different approach to the 
HAM which is based on the Least Action Principle (LAP). For Lagrangian 
systems, 
the use of LAP leads to 
different 
results with respect to the standard approach based on the residual error square. 
This approach, to which we refer to as Lagrangian Homotopy Analysis Method (LHAM) might lead to 
improved efficacy and to a reduction of computing time in particular for highly nonlinear problems. 

In general, the optimization principles, like LAP, play a fundamental role in many areas of natural sciences \cite{Lanczos1970, PhilippeChoquard1992, Coopersmith.2017, MichelBierlaire2018}. In a broad mathematical sense, the goal of the optimization principle is to maximize or to minimize a function by selecting the best option from a set of allowed ones. A simple example is the way light rays travel between two media, where the function to optimize is the time needed to go from point $A$ to point $B$: as a result light's trajectory from $A$ to $B$ is not straight. In classical and quantum mechanics and in field theory the function to optimize is the {\it action} \cite{Feynman2006}. For instance, in classical mechanics, the action $S$ is a functional of different trajectories with given initial and final states, and the actual trajectory is the one around which the action is stationary, $\delta S = 0$. An important point to be remarked is that if the Hamiltonian is a convex function of the canonical variables, the classical trajectory is 
a {\it minimum} of the action \cite{PhilippeChoquard1992}. However, in general, the trajectory  which optimizes the action could also be a maximum or even (for Lagrangian systems with several degrees of freedom) a saddle \cite{Mackay1984, Mackay1989, Meiss2015, Shpielberg2016, Shpielberg2017}. 

In order to implement the LAP
and optimize the convergence of the solutions given by the HAM, we develop an approach consisting in finding the optimal value of the HAM parameter $c_0$ for which the approximated action $S$ obtained by HAM is stationary. We will show that indeed one can generally find many values of $c_0$ that fulfill this optimization principle. Therefore, in order to choose the best value of $c_0$ among those selected by this principle, we have to introduce an additional criterion. Using the fact that energy is conserved and known from the initial condition, we have 
used a 
``best energy conservation criterion" in order to choose the optimal value of $c_0$. We will argue that it is less convenient 
to minimize the (modulus of) the difference 
$\Delta E$ between the energy of the approximate solution and the initial, exact one. The reason is that one finds several solutions for $c_0$ minimizing $\Delta E$ and one {\it a priori} does not know what to choose. The point is that the HAM method when applied to Hamiltonian systems determines the approximations to the solution of the problem without using energy conservation -- actually, the energy is fixed by the intial conditions and indeed $\Delta E$ can be either positive or negative when plotted as a function of $c_0$.
%

We have tested the proposed approach in four case studies: ({\it i}) the harmonic oscillator (for which the exact solution is known), ({\it ii}) the quartic anharmonic oscillator, {\it (iii)} the cubic anharmonic oscillator 
 and ({\it iv}) the Korteweg-de Vries (KdV) partial differential equation \cite{Zakharov1971, Davydov1980}. 
The paper is organized as follows. In section~\ref{method}, we first give a 
reminder on the HAM and then we present our approach based on LAP. In section~\ref{apply}, we apply our method to: the harmonic oscillator (\ref{harmonic Oscillator}), the quartic anharmonic oscillator (\ref{quarticOscillatorySysytem}), 
the cubic anharmonic oscillator (\ref{cubicOscillatorySysytem}), and the KdV equation (\ref{KdV equation}). Concluding remarks and perspectives are given in section~\ref{conclusions}. The paper concludes with three appendices.

\section{Outline of the method}
\label{method}

\subsection{The 
Homotopy Analysis Method}
\label{homotopydescription}

A 
description of the HAM is outlined in this section, referring for simplicity to ordinary differential equations. For this purpose, let us consider the following general nonlinear ordinary differential equation

\begin{equation}
\mathcal{N}\left[x\left(t\right)\right]=0
 \label{HAM1},
\end{equation} 
where $\mathcal{N}$ is a nonlinear operator, $t$ denotes an independent variable, and $x\left(t\right)$ is an unknown function to be determined, respectively. Eq.~(\ref{HAM1}) has to be supplemented by the associated boundary conditions, e.g., for a first order differential equation by the value of $x(t=0)$. Of course, $x(t)$ can be as well a multi-component vector.

In HAM one makes use of 
homotopy, a basic concept in topology, by writing \cite{SJLiao2003}:

\begin{equation}
\left(1 -q\right)\mathcal{L}\left[ \phi\left(t , q\right)  \,-\, x_0\left(t\right)\right]=c_0 \,q \,\mathcal{H}\left(t \right)\,\mathcal{N}\left[\phi\left(t,q\right)\right]
\label{HAM2},
\end{equation} where $q\in \left[0,1\right] $ is the  embedding parameter 
called the homotopy embedding parameter, $c_0$ is a non-zero auxiliary parameter that we call the HAM parameter, 
$\mathcal{L}$ is an auxiliary linear operator with the property ${\mathcal{L}\left[ x_0 \right]=0}$. Moreover, 
$\mathcal{N}$ is the nonlinear operator related to Eq.~(\ref{HAM1}), $x_0\left(t\right)$ is an initial guess for $x\left(t\right)$, $\mathcal{H}\left(t \right)$ is an auxiliary function to adjust the sought solution, and $\phi\left(t , q\right)$ is the solution of  Eq.~(\ref{HAM2}) for $q\in \left[0,1\right] $, respectively. Notice that, in the frame of the HAM, we have 
freedom to chose the auxiliary linear operator $\mathcal{L}$, the initial guess $x_0\left(t\right)$, the auxiliary function  $\mathcal{H}\left(t \right)$, and the value of the HAM parameter $c_0$. 

When $q=0$\, due to the property $\mathcal{L}\left[ x_0 \right]=0$, we get from Eq.~(\ref{HAM2}) the solution

\begin{equation}
\phi\left(t , 0\right) = x_0\left(t\right)
 \label{HAM4}.
\end{equation}
When  $q=1$, since $c_0\ne 0$ \, and \, $\mathcal{H}\left(t \right)\ne 0$ almost everywhere, Eq.~(\ref{HAM2}) is equivalent to the original nonlinear equation (\ref{HAM1}) so that we get

\begin{equation}
\phi\left(t , 1\right) = x\left(t\right)
 \label{HAMM5}.
\end{equation}
Eq.~(\ref{HAM2}) is usually referred to as the {\it zeroth-order} deformation equation. 
Expanding $\phi\left(t , q\right)$ in Maclaurin series with respect to $q$ at $q=0$, one obtains

\begin{equation}
\phi\left(t , 1\right) = x_0\left(t\right)\,\,+\,\,\sum\limits_{m=  1}^{+\infty } { x_m\left(t\right)\, q^m }  
 \label{HAM5},
\end{equation} 
where the series coefficients $x_m$ are defined by

\begin{equation}
x_m\left(t\right)=\frac{1}{m!}\frac{\partial^m \phi\left(t , q\right)}{\partial q^m}{\Biggl{|}}_{q=0}
 \label{HAM6}.
\end{equation}

If the auxiliary linear operator $\mathcal{L}$, the initial guess $x_0\left(t\right)$, the HAM parameter $c_0$ and the auxiliary function $\mathcal{H}\left(t \right)$ are 
properly chosen, the homotopy series (\ref{HAM5}) converges at $q=1$, then using the relationship $\phi\left(t , 1\right)=x\left(t\right)$, one has the so-called homotopy series solution

\begin{equation}
x\left(t\right)= x_0\left(t\right)\,\,+\,\,\sum\limits_{m=  1}^{+\infty } { x_m\left(t\right) }  
 \label{HAM7},
\end{equation} 
which must be one of the solutions of the original nonlinear equation (\ref{HAM1}) 
\cite{SJLiao2003, SJLiao2014}. Substituting the series (\ref{HAM5}) into the zeroth-order deformation (\ref{HAM2}) and equating the like-power of $q$, we get the high-order approximation equations for $x_m\left(t\right)$, also called the $m$th-order deformation equations 

\begin{equation}
\mathcal{L}\left[  x_m\left(t\right)  -\chi_m  \, x_{m-1}\left(t\right)\right]=c_0\, \mathcal{H} \left(t \right)\mathcal{R}_{m-1}\left[  x_{m-1}\left(t\right)   \right]
\label{HAM8},
\end{equation} 
where $\mathcal{R}_{k\equiv m-1}$ is the so-called  $k${\it th}-order homotopy derivative operator given by

\begin{equation}
\mathcal{R}_{k} \left[\phi\left(t\right)\right]=\frac{1}{k!}\frac{\partial^k \, \mathcal{N}\left[  \phi\left(t\right)  \right] }{\partial q^k}{\Biggl{|}}_{q=0}
 \label{HAM9},
\end{equation} 
and we defined 

\begin{equation}
\chi_m  = \left\{ {\begin{array}{*{20}c}
   0, & {m \le 1}  \\
   1, & \textrm{elsewhere}.  \\
\end{array}} \right.
 \label{HAM10}.
\end{equation}
Notice that, the right-hand side of term $\mathcal{R}_{m-1}$ in Eq.~(\ref{HAM8}) is only dependent upon 

\begin{center}
$x_0\left(t\right), \,\, x_{1}\left(t\right), \,\,x_{2}\left(t\right),\,\, ...,\,\, x_{m-1}\left(t\right)$,
\end{center} 
which are known for the $m$th-order deformation equation described above. Finally, an $M${\it th}-order approximate analytic solution of practical interest is given by truncating the homotopy series (\ref{HAM7}) up to $M$. The exact solution is given by the limit 

\begin{equation}
x\left(t\right)\,\,=\,\,\lim_{M\to\infty} x_M\left(t\right)
 \label{HAM12}.
\end{equation}

At the $m${\it th}-order approximation, the value of the HAM parameter $c_0$ can be determined by the minimum of the residual error square $\epsilon_m$ of the original governing equation. $c_0$ corresponds to the minimum of the residual error square. i.e., 

\begin{equation}
\frac{d \, \epsilon_m\left(c_0\right)}{d c_0}=0
 \label{HAM13},
\end{equation} 
with

\begin{equation}
\epsilon_m\left(c_0\right)\,=\, \int_{\Omega}\left({\mathcal{N}\left[  \sum\limits_{k=  0}^{m } { x_k\left(r\right) }    \right]}\right)^2 d r,
 \label{HAM14}
\end{equation}
$\Omega$ being the domain of interest for the problem under consideration. 

To simplify the computation, if it is known that the integrand ${\mathcal{N}\left(\sum\limits_{k=  0}^{m } { x_k\left(r\right) }\right)}$ is positive, then it is 
convenient to use the 
residual error

\begin{equation}
\epsilon_m\left(c_0\right)\,=\, \int_{\Omega}\left({\mathcal{N}\left[  \sum\limits_{k=  0}^{m } { x_k\left(r\right) }    \right]}\right) d r
\label{HAMM14}.
\end{equation} 
Obviously, the more quickly $\epsilon_m\left(c_0\right)$ in Eq.~(\ref{HAM14}) or Eq.~(\ref{HAMM14}) decreases to zero, the faster the corresponding homotopy series solution (\ref{HAM7}) converges and the accuracy of the homotopy approximations increases. At the 
$M$th-order of approximation, the 
value of the HAM parameter $c_0$ is given by the minimum of  $\epsilon_M\left(c_0\right)$, corresponding generally to a nonlinear algebraic equation to be solved from Eq.~(\ref{HAM13}).

The HAM depends upon the number of the HAM parameters $c_0$, but it is in general time-consuming to find out the HAM parameter, especially at high-order of approximations for complicated nonlinear problems. When there are more than one unknown parameters, the needed time considerably increases 
so that the exact residual error square 
can be difficult to use in practice. Therefore, it can be relevant both conceptually and for practical applications to find ways to decrease the computation times and/or to obtain 
more accurate results. 

In this logic we consider in the following Lagrangian systems, for which Eq.~(\ref{HAM1}) is the equation of motion and an action $S$ can be defined. As well known, from the extremization of the action, one gets through the Euler-Lagrange method the equation of motion \cite{PhilippeChoquard1992}. For the sake of simplicity, in this paper we restrict to those Lagrangian systems for which one can pass to the Hamiltonian description, and then define the energy as constant of motion.  
Therefore the point we address in the next section is if we can take advantage of the geometrical structure of Lagrangian systems by making use of the 
LAP.

\subsection{
Lagrangian Homotopy Analysis Method using the Least Action Principle}
\label{Lhomotopydescription}

To set the notation, we remind that for a Lagrangian system with coordinates 
$x=x(t)$ and Lagrangian $L=L(x,\dot{x})$, from the LAP and the extremization of the action $S=\int_0^t L(x(\tau),\dot{x}(\tau)) d\tau$ one gets the Euler-Lagrange equations 
\begin{equation}
\frac{d}{dt} 
\frac{\partial L}{\partial \dot{x}} = \frac{\partial L}{\partial x},
\label{EL}
\end{equation}
where $x$ may also refer to several coordinates $x_i$'s \cite{Goldstein1980}. By performing a Legendre transform, one can then construct the Hamiltonian $H=H(x,p)$. We will restrict ourself to cases in which both $L$ and $H$ do not explicitly depend on time. It is intended that the Euler-Lagrange equations~(\ref{EL}) when written for $x(t)$ is just Eq.~(\ref{HAM1}) [that is the reason for which we denote the Lagrangian coordinate $x(t)$ and not $q(t)$ as it is also customary]. 
One can also consider the continuum limit, 
where the coordinates $x_i(t)$'s depend on a continuous parameter $X$: $x_i(t) \to x(X,t)$: we do not write here the corresponding formulas, referring to \cite{Goldstein1980}.




Let us now introduce the problem of using the LAP to determine the HAM parameters. The HAM, when truncated to the order $M$, produce approximate solutions of the equation of motions. Such approximate orbits depend on the HAM parameter $c_0$ and one is lead to the question of how to use the Lagrangian structure of the original problem. This is an instance of having an approximate solution, depending on a single or more parameters, for the equations of motion of a problem (linear or not), and then find the best among them, i.e., the closest -- in some sense to be defined -- to the exact, unknown solution of the problem at hand.

As first, one would think to minimize the energy to get the HAM parameter $c_0$. However, we remind that the initial condition $x(t=0)$ is known and then the energy $E=H(x(0),p(0))$ itself is in turn exactly known. One can then calculate the energy of the approximate solution at a certain time $t$. Denoting such an energy by $E_{approximate}(t)$ and the exact energy $E$ by $E_{exact}$, it could be that $E_{approximate}(t)$ is larger or smaller than $E_{exact}$. Then one should minimize the modulus of the difference $|E_{exact}-E_{approximate}(t)|$ to find the optimal value of $c_0$ at that time $t$. 

However, when carrying out this procedure also for simple problems such as the harmonic oscillator (where the exact solution is readily determined), one realizes that $|E_{exact}-E_{approximate}(t)|$ plotted as a function of $c_0$ (at the fixed time $t$) is rather flat, and the determination of the optimal $c_0$ difficult. 

To circumvent this problem, we can invoke and use the LAP. A fundamental property of the LAP is that for any first-order variation away from the optimal path, the change in time is zero, i.e., the trajectory is such that the corresponding action has an extremum. Therefore, we choose -- at a certain time -- to extremize by the LAP the approximated action $S_{approximate}$, which depends on $c_0$: 

\begin{equation}
{\frac{\partial \,  S_{approximate}\left(c_0\right)}{\partial c_0}}{\Biggl{|}}_{c_0}=0
 \label{homotopic3}.
\end{equation} 
In the following, to not make heavier the notation, we do not write explicitly $S_{approximate}\left(c_0\right)$, but it is intended that the action is calculated by the HAM approximate solution $x_{approximate}(c_0, t)$.

Eq.~(\ref{homotopic3}) has more than a solution, actually a set of solutions. The cardinality of such a set increases  with the order of the approximation. 
To select the optimal $c_0$ we then use the conservation of the energy, and the chosen HAM parameter is the one that gives the best conservation of the energy of the system at a finite time, and consequently, the better convergence of the HAM approximate solution to the exact one at that given time. 

This scheme will reveal very useful for nonlinear differential equations, especially for higher-order nonlinear problems, where it appears that it is 
more difficult (and time consuming) to find out $c_0$ using the standard residual error square method. To start with, we illustrate the method for the simplest case of an harmonic oscillator.








\section{Applications of the Lagrangian Homotopy Analysis Method}
\label{apply}

\subsection{Harmonic oscillator}
\label{harmonic Oscillator}


Linear and  nonlinear differential equations appear in many fields and play a crucial role in modeling complex physical phenomena such as vibrations in lattice and pulse shapes in biological network systems, e.g., see \cite{Andrianov2020, Apaza2020, Holfmann2011, Coveney2005}. Here, we start the section by considering the classical harmonic oscillator system defined by the following Hamiltonian:  

\begin{equation}
H\left(x, p \right)=\frac{p^2}{2{\cal M}} \,\,+ \,\, \frac{1}{2}k\,x^2  
 \label{App1},
\end{equation} 
where ${\cal M}$ is the mass of the oscillator and $k$ the spring constant. The corresponding equation of motion 
reads

\begin{equation}
x_{tt} \,\,+\,\,\, \omega^2 \,x =0  \,\,\,\,\,\,\,\,\,\,\,\,\,\,\,\,\,\,\,\,\,\,\,\,\,\, \textrm{with} \,\,\,\,\,\,\,\,\,\,\,\,\, \omega=\sqrt{\frac{k}{{\cal M}}} .
 \label{App2}
\end{equation}
With the initial conditions 
\begin{center}
$x\left( 0 \right)=1$\,\,\,\,\,\,\textrm{and} \,\,\,\,\,\, $x_t\left( 0 \right)=0$,
\end{center} 
Eq.~(\ref{App2}) 
possesses the exact solution $$x\left( t \right)=\cos\left( \omega \, t\right).$$

The approximate solution of Eq.~(\ref{App2}) is obtained by means of the HAM using the following linear operator, the initial guess and auxiliary function:

\begin{center}
{$ \mathcal{L} \equiv \frac{d^2}{d t^2}$},\,\,\,\,\,\,\,\, $x\left( 0 \right)=1$, \,\,\,\,\,and \,\,\,\,\,  $\mathcal{H}\left( t \right)=1$,
\end{center} 
respectively. Notice that we are denoting the HAM linear operator and auxiliary function by $\mathcal{L}$ and $\mathcal{H}$, to be not confused with the Lagrangian $L$ and the Hamiltonian $H$. Since the linear operator should be chosen in order to permit the initial approximation taken, i.e., 

\begin{equation}
{\mathcal{L}\left[x_0\left( t \right)\right] \equiv \frac{d^2\, x\left( 0 \right)}{d t^2}=0}
 \label{apApp1},
\end{equation} 
we clearly see that Eq.~(\ref{apApp1}) holds, i.e.,  at $q=0$, the linear operator satisfied the zeroth-order deformation. Feeding the initial guess $x_0\left( t \right)=1$ into Eq.~(\ref{HAM8}) where $$\mathcal{R}_{m-1}\left(x_{m-1}\left(t\right)\right)=x_{{(m-1)}tt} \,\,+ \,\,x_{m-1},$$ we get for $m=1,\cdots,4$

\begin{equation}
x_1\left(t\right)\,\,=\,\, \frac{c_0}{2}\,\omega^2t^2
 \label{App3},
\end{equation}

\begin{equation}
x_2\left(t\right)\,\,=\,\,\,\,x_1\left( t \right)\,\, +\,\, \frac{c_0^2}{2}\,\omega^2t^2 \,\,+\,\,\frac{c_0^2}{24}\,\omega^4t^4 
 \label{App4},
\end{equation}

\begin{equation}
x_3\left(t\right)\,\,=\,\, x_2\left(t\right) +  \frac{c_0^2}{2}\,\omega^2t^2  \,\, +\,\,  \frac{c_0^3}{2}\,\omega^2t^2   \,\, +\,\,  \frac{c_0^3}{24}\,\omega^4t^4  \,\, +\,\,  \frac{c_0^2}{24}\,\omega^4t^4 \,\, +\,\,  \frac{c_0^3}{24}\,\omega^4t^4 \,\, +\,\,  \frac{c_0^3}{720}\,\omega^6t^6
 \label{App5},
\end{equation}

\begin{equation}
x_4\left(t\right)\,\,=\,\, x_3\left(t\right) \,+ \, \frac{c_0^2\,\omega^2}{5040}\left[ 2520\left(1+c_0\right)^2t^2 \,+\,210\left(1+4c_0+3c_0^2\right)\omega^2 t^4 \,+\, 7c_0\left(2+3c_0\right)\omega^4t^6 \,+\, \frac{1}{8} c_0^2\omega^6t^8  \right]
 \label{App05}.
\end{equation}

Several additional expressions can be found using mathematical software. 
Approximation series solution can be obtained to any desired number of terms. We can therefore set up an approximate solution through the following series expansion:

\begin{equation}
x\left(t\right)\,\,=\,\, x_0\left(t\right)  \,\, +\,\, x_1\left(t\right)\,\, +\,\, x_2\left(t\right)\,\, +\,\, x_3\left(t\right) \,\, +\,\, x_4\left(t\right) + \,\,...,
 \label{App6}
\end{equation} i.e.,

\begin{equation}
\begin{array}{l}
x\left(t\right)\,\,=1+\left( 2c_0 \omega^2 + 3 c_0^2 \omega^2 + 2 c_0^3 \omega^2 + \frac{c_0^4\omega^2}{2} \right)t^2+\frac{1}{24}\left( 6 c_0^2 \omega^4 + 8 c_0^3 \omega^4 + 3 c_0^4 \omega^4   \right)t^4  
\\
\hspace{1.4cm}+\left( \frac{c_0^3\omega^6}{180}\frac{c_0^4\omega^6}{240}   \right)t^6+\frac{c_0^4\omega^8}{40320}t^8\,+ \,\,0\left[t\right]^9.
\end{array}
\label{App06}
\end{equation}

With $c_0\,=-1$, Eq.~(\ref{App06}) reads

\begin{equation}
x\left(t\right)\,\,=\,\, 1  \,\, - \,\, \frac{\omega^2}{2}t^2\,\, +\,\, \frac{\omega^4}{24}t^4\,\, - \,\, \frac{\omega^6}{720}t^6\,\,+ \frac{\omega^8}{40320}t^8\, + \,\,0\left[t\right]^9
 \label{App07},
\end{equation} 
which can be rewritten as follows

\begin{equation}
x\left(t\right)\,\,=\,\, 1  \,\, - \,\, \frac{\omega^2}{2!}t^2\,\, +\,\, \frac{\omega^4}{4!}t^4\,\, - \,\, \frac{\omega^6}{6!}t^6\,\,+ \frac{\omega^8}{8!}t^8\,\,+ \,\,0\left[t\right]^9.
\label{App7}
\end{equation}
The homotopy series solution then reads

\begin{equation}
x\left(t\right)\,\,=\,\, \sum\limits_{m= 0}^{+\infty } { \frac{\left(  -1 \right)^m \left(  \omega \,t \right)^{2m} }{\left(  2 m \right)!} } \,\,=\,\, \cos\left(  \omega \, t \right)
 \label{App8},
\end{equation} 
showing that the exact solution is found for $c_0=-1$.

Let us see whether one can find $c_0$ by making use of the LHAM. 
Using the LAP and plotting the action as a function of $c_0$ at a given time, we observe a rather flat behaviour. Zooming in on the flat region we observe a set of minima and maxima that correspond to the set of HAM parameters $c_0$ that we look for. 
Notice that we select minima and maxima even though we know that the Hamiltonian $H$ is convex in $x$ and $p$ and therefore one knows that the classical trajectory is 
a minimum of the action \cite{PhilippeChoquard1992}. However, we select also maxima because this result is valid for the exact action, but here we are extremizing the approximated one.

The energy of the harmonic oscillator system (\ref{App1}) is

\begin{equation}
H = \frac{1}{2}\left[  {\cal M} \left(\frac{d\, x\left(t\right)}{dt}\right)^2   \,\,+ \,\,  k\, x^2\left(t\right) \right]
 \label{App10},
\end{equation} 
and the action reads

\begin{equation}
S= \int_{0}^{t}{\left(\frac{{\cal M}}{2}\left(\frac{d\, x\left(t\right)}{dt}\right)^2   \,\,- \,\,  \frac{k}{2}\, x^2\left(t\right) \right)dt}
\label{harmonicaction1}.
\end{equation}

\begin{figure}[ht]
\centering
\includegraphics[width=2.3in,angle=0]{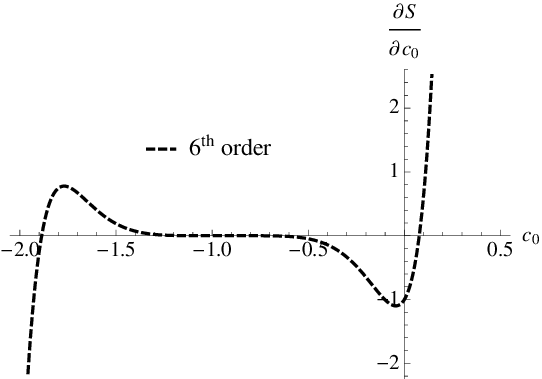}(a)\,\,\,\,\,\,\includegraphics[width=2.3in,angle=0]{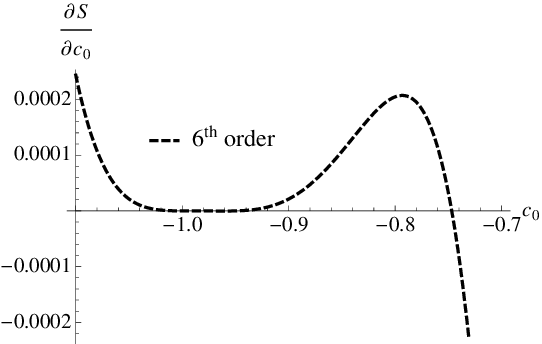}(b)
\includegraphics[width=2.3in,angle=0]{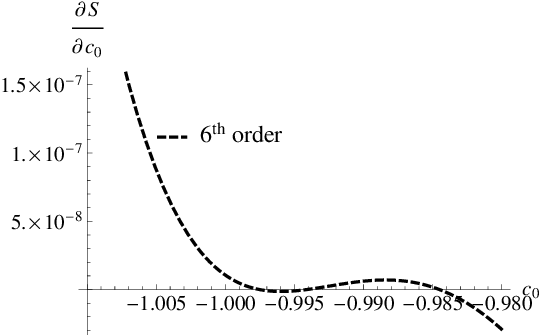}(c)\,\,\,\,\,\,\includegraphics[width=2.3in,angle=0]{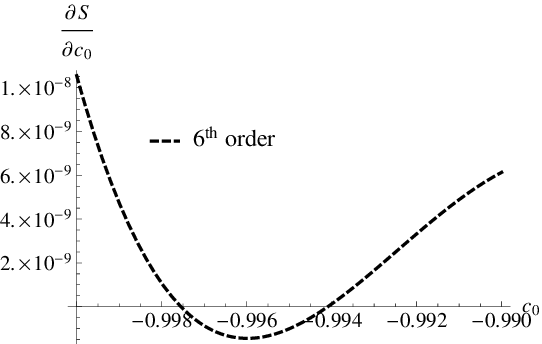}(d)
\caption{The derivative of the action with respect to $c_0$, $\frac{\partial \,S}{\partial \,c_0}$, at the $6th$-order HAM approximation using LAP for the harmonic oscillator (\ref{App2}) with $t=1$ (and $\omega=1$). In the panels (b)-(c)-(d), we progressively zoom into the flat region of panel (a) in order to look for minima and maxima.} 
\label{fig:harmonicsolutions}
\end{figure}

Extremizing the action, ${\frac{\partial \,  S\left(c_0\right)}{\partial c_0}}{\Bigl{|}}_{c_0}=0$, we observe a flat region which contains a set of HAM parameters $c_0$, as seen in Fig.~\ref{fig:harmonicsolutions} and Table~\ref{tab:6thhomotopicquadratic} at the $6^{th}$-order approximation.

\begin{table}[h!t]
\begin{tabular}{|p{1.5cm}|p{2.5cm}|p{3.cm}|}
\hline
\multicolumn{3}{|c|}{Set of HAM parameters $c_0$ extremizing the approximated action at the $6^{th}$-order 
} \\
 \hline
$c_0$  & $E\left(t=1, \,\,c_0 \right)$ & $\left| E_{exact} - E_{approximate}\right|$ \\
 \hline
-1.8820               &0.8589                            & 0.3589\\
 \hline
-0.9977              &0.500000000008              &8.91$\times10^{-12}$ \\
 \hline
-0.9940              &0.49999999996             & 3.4$\times10^{-11}$\\
 \hline
-0.9846              &0.50000000051            & 5.1$\times10^{-10}$\\
 \hline
-0.9490              &0.49999993                 & 6.7$\times10^{-8}$\\
 \hline
-0.7475             &0.5002                          &0.0002\\
 \hline
0.0810              & 1.0397                         & 0.5397\\
  \hline
\end{tabular}  
\caption{The set of HAM parameters, $c_0$, at the $6$th-order approximation for the harmonic oscillator (column 1). In column 2 we report the energy at time $t=1$ (and $\omega=1$) corresponding to each value of the corresponding HAM parameter $c_0$ (in this table and the following, the error is estimated to be $(1)$ at the last reported digit, e.g., $c_0=-1.8820$ stands for $c_0=-1.8820(1)$. Columns 3 shows the variation of the energy using the optimization of the action. We observe that the value of $c_0$ which allows the best conservation of energy at $t=1$ is $c_0= -0.9977$. This is therefore the value we select at the $6$th-order approximation.}
\label{tab:6thhomotopicquadratic}
\end{table}

In Table~\ref{tab:harmonicccpaction}, we see that by increasing the order of approximation, the HAM parameter $c_0$ selected as discussed converges to its exact value $-1$ and the corresponding energy of the system to its exact value $E=0.5$.

\begin{table}[h!t]
\begin{tabular}{ |p{1.5cm}|p{1.5cm}|p{2.9cm}|p{2.8cm}|}
\hline
\multicolumn{4}{|c|}{Selected HAM parameter $c_0$ at different orders of approximation} \\
 \hline
$m$th-order & \centering{$c_0$}  & \centering{$E\left(t=1, \,\,c_0 \right)$} &$\left| E_{exact} - E_{approximate}\right| $ \\
 \hline
 $1^{st}$       &     0.5945              &1.0181                           & 0.5181 \\
  \hline
 $2^{nd}$      &    -0.9477             &0.50433                        &$4.3\times10^{-3}$\\ 
  \hline
 $3^{rd}$      &     -0.9804            &0.499919                       &$8.0\times10^{-5}$\\
  \hline  
 $4^{th}$      &    -0.9925             &0.5000005                     &$5.08\times10^{-7}$ \\
  \hline
 $5^{th}$      &    -0.9959             &0.4999999973              &$2.69\times10^{-9}$  \\
  \hline 
 $6^{th}$      &    -0.9977             &0.500000000008         &8.91$\times10^{-12}$ \\   
\hline
 $7^{th}$   & -0.9984                  &0.49999999999997       &$2.21\times10^{-14}$ \\
 \hline
 $8^{th}$   & -0.9999                 &0.499999999999998      & $1.16\times10^{-15}$     \\
  \hline
\end{tabular}  
\caption{The HAM parameter ${c}_0$ (column 2) with increasing order of the approximation (column 1) for the HAM based on the optimization of the action. In column 3, we simply report the HAM solution at time $t=1$. Column 4 shows the error of the energy using our approach. $\omega$ and the time $t$ are taken as one. 
}
\label{tab:harmonicccpaction}
\end{table} 

\begin{figure}[ht]
\centering
\includegraphics[width=2.3in,angle=0]{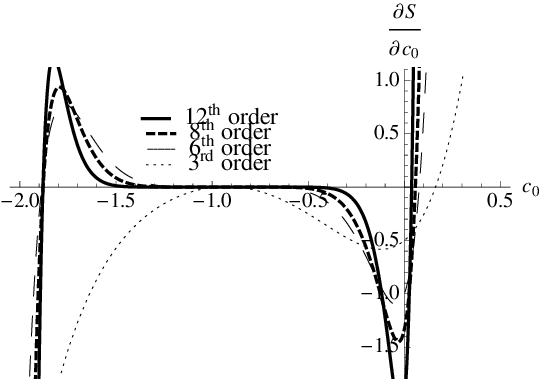}(a)\,\,\,\,\,\,\includegraphics[width=2.3in,angle=0]{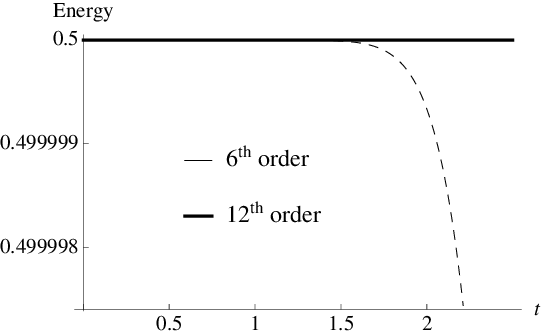}(b)
\includegraphics[width=2.3in,angle=0]{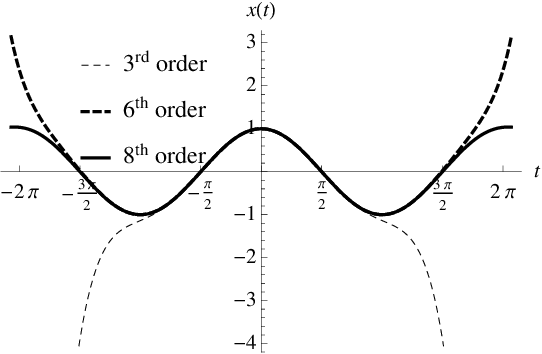}(c)\,\,\,\,\,\,\includegraphics[width=2.3in,angle=0]{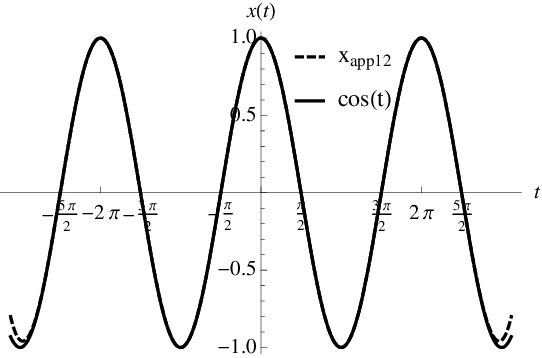}(d)\\
\caption{Panel (a) shows the derivative of the action with respect to $c_0$, $\frac{\partial \,S}{\partial \,c_0}$, at different orders of HAM approximation with LAP with $t=1$ (and $\omega=1$). Panel (b) reports the energy vs. time showing 
energy conservation up to a time $t$ which increases with the orders of HAM approximation. 
Panel (c) shows the HAM approximation of the solution at increasing order. In panel (d) we compare the exact solution with the $12^{th}$-order of approximation. 
Plotting the difference between the exact solution and the approximate one at the $12^{th}$-order of the approximation, one sees that for $t$ between $-\pi$ and $\pi$ the error is smaller than $10^{-12}$. $\omega$ is taken as one.} 
\label{fig:harmonsolutions}
\end{figure}

Looking at Fig.~\ref{fig:harmonsolutions}, we see that at higher orders of HAM approximation, the energy of the harmonic system is conserved, and consequently, we reach to the exact solution $x\left(t\right)=\cos\left(\omega t\right)$. It is straightforward to observe that the HAM approximate solution will be more accurate when the order of the approximation tends to larger value of the order of the approximation. This provides a benchmark of the reliability of the proposed approach.

\begin{figure}[ht]
\centering
\includegraphics[width=1.5in,angle=0]{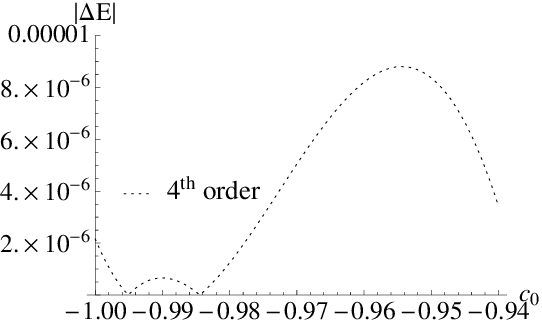}(a)\includegraphics[width=1.5in,angle=0]{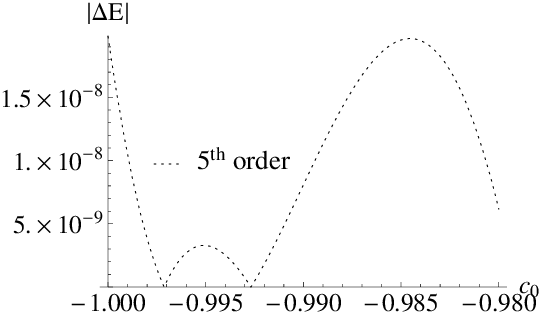}(b)
\includegraphics[width=1.5in,angle=0]{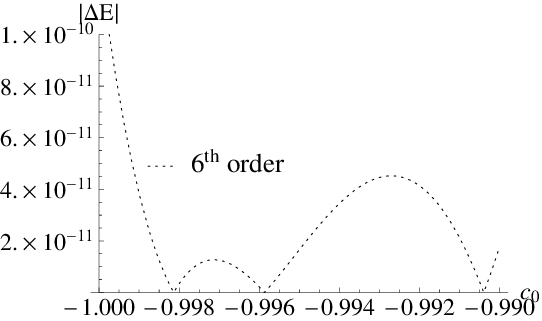}(c)\\
\includegraphics[width=1.5in,angle=0]{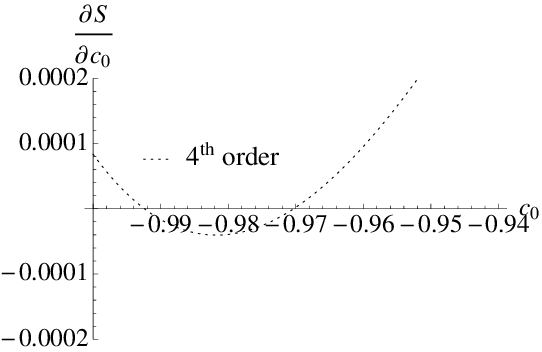}(d)\includegraphics[width=1.5in,angle=0]{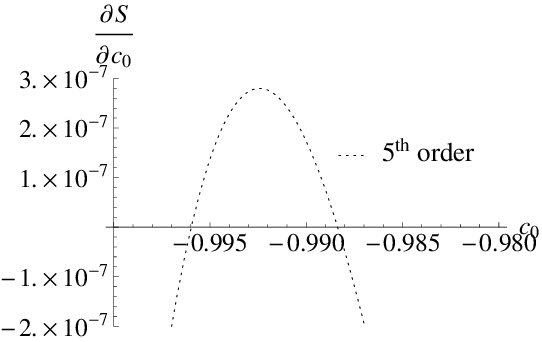}(e)
\includegraphics[width=1.5in,angle=0]{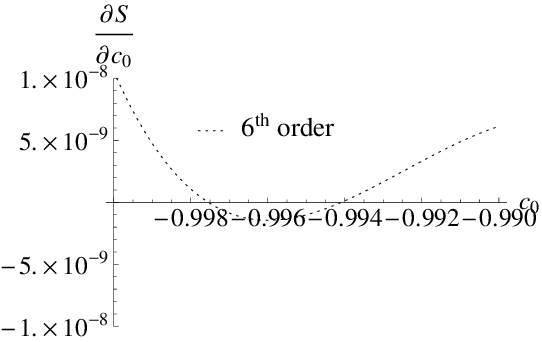}(f) 
\caption{Panels (a), (b) and (c) show  $|\Delta E|$ as a function of $c_0$ at different orders of HAM approximation. 
Panels (d), (e) and (f) show the derivative of the action with respect to $c_0$, $\frac{\partial \,S}{\partial \,c_0}$, at different orders of HAM approximation with LAP, respectively. $\omega$ and the time $t$ are taken as one.} 
\label{fig:harmonicEnergyflatness}
\end{figure}

We pause here to comment about the usefulness of extremizing the action before choosing the value of $c_0$ for which $|E_{exact}-E_{approximate}| \equiv 
|\Delta E|$ is minimum. Indeed, if one calculates at the $M$-th order directly $|\Delta E|$ as a function of $c_0$ for a given time, then it emerges that also in this very simple case of the harmonic oscillator one finds that 
$\Delta E$ can be positive or negative (notice that $E_{exact}$ is known from initial conditions) and there are several values of $c_0$ for which $|\Delta E|$ is minimum. So one does not know {\it a priori} what among these values to choose. Finally, we observe that $\Delta E$ is rather flat increasing the order of the approximation, flatter than $\partial S/\partial c_0$, and it numerically not straightforward to estimate the set of values of $c_0$ for which $|\Delta E|$ is vanishing. These features are illustrated in Fig.~\ref{fig:harmonicEnergyflatness}.

Let us move forward by introducing a nonlinear term in the system. In general, by increasing the nonlinearity we also increase the difficulty in solving the nonlinear equation. As case studies, we solve the {\it quartic}- and {\it cubic}- nonlinear anharmonic problems, respectively  
and test the proposed approach.

\subsection{Quartic anharmonic oscillator}
\label{quarticOscillatorySysytem}

Nonlinear equations are much more difficult to solve than linear ones, especially by means of analytic methods. Thus, it is not guaranteed that one can 
always 
obtain approximate solutions for any given nonlinear problem. 
Here and in the next section we investigate whether the HAM combined to LAP scheme 
provide a good alternative. To start with, 
let us consider here a one-dimensional quartic anharmonic oscillators problem with a repulsive $(\gamma>0)$ quartic potential. For $\gamma<0$, the potential is attractive and the system is unstable near $x=0$. 
The corresponding Hamiltonian reads:

\begin{equation}
 H\left(x, \, p \right)=\frac{p^2}{2} \,\,+ \,\, \frac{k}{2}\,x^2  \,\,- \,\, \frac{\gamma}{4}\,x^4
 \label{Anh1},
\end{equation} 
with the mass ${\cal M}$ of the system being taken here as ${\cal M}=1$. The corresponding equation of motion reads: 

\begin{equation}
x_{tt} \,\,+\,\, k\,x \,\,-\,\, \gamma\, x^3 =0 .
 \label{Anh2}
\end{equation} 
With initial conditions given by 

\begin{center}
$x\left( 0 \right)=0$,\,\,\,\,\,\,\,\,\,\,\,\,\, and \,\,\,\,\,\,\,\,\,\,\,\,\,  $x_t\left( 0 \right)=\frac{1}{\sqrt{2}}$,
\end{center}
the exact solution reads

\begin{equation}
x\left( t \right)=\tan\left( \frac{1}{\sqrt{2}}t\right),\,\,\,\,\,\,\,\,\, \textrm{with}\,\,\,k=\gamma=1
 \label{Anh3}.
\end{equation}
Let us use our proposed approach to solve Eq.~(\ref{Anh2}). The solution reads 

\begin{center}
$x\left(t\right)= x_0\left(t\right)\,\,+\,\,\sum\limits_{m=  1}^{M } { x_m\left(t\right) }$,
\end{center} 
where

\begin{equation}
x_m\left( t \right)\,\,=\,\chi_m\,\,x_{m-1}\left( t \right)\,\,+\,\, c_0 \, \mathcal{L}^{-1}\left[  \mathcal{R}\left( x_{m-1}   \right)   \right],
 \label{Anh4}
\end{equation} and 

\begin{equation}
\mathcal{R}\left( x_{m-1}   \right) \,\,=\,x_{{(m-1)}tt} \,\,+\,\, x_{m-1}\,\,-\,\,\sum\limits_{i=  0}^{m-1 } { x_{m-1-i} }\sum\limits_{j=  0}^{i } { x_j\,x_{i-j}}. 
\label{Anh5}
\end{equation}

We choose the following initial guess 

\begin{equation}
x_0\left( t \right)\,\,=\,a\,t
 \label{Anh6},
\end{equation} 
where $a$ is a constant to be later determined. Since the linear operator $\mathcal{L}$ should be chosen in order to allows for the initial approximation,  $\mathcal{L}\left[  x \right]\,\equiv\, x_{tt}$ is a good choice. The linear operator satisfies the zeroth-order deformation equation:

\begin{equation}
\mathcal{L}\left[  x_0 \right]\,\equiv\, \frac{d^2\,( a\,t) }{d t^2}=0 .
 \label{Anh8}
\end{equation}
Choosing the auxiliary parameter $\mathcal{H}\left( t \right)=1$ and feeding the initial guess Eq.~(\ref{Anh6}) into Eq.~(\ref{Anh2}) with $k=\gamma=1$, we get for $m=1,\cdots,4$

\begin{equation}
x_1\left( t \right)\,\,=\,-\,\frac{c_0\, a}{4}\left( -\frac{2}{3}t^3 \,\,+\,\, \frac{a^2}{5}t^5    \right)    
 \label{Anh9},
\end{equation}

\begin{equation}
x_2\left( t \right)\,\,=\,\,x_1\left( t \right)\,\,+\,\,\frac{c_0^2\, a}{480}\left(  80 t^3\,-\,4( -1 + 6 a^2 )t^5\,-\, \frac{44}{7}a^2t^7 + a^4t^9    \right)
 \label{Anh10},
\end{equation}

\begin{equation}
\begin{array}{l}
x_3\left( t \right)\,\,=\,\,x_2\left( t \right)\,\,+\,\,\frac{c_0^2\, a}{16800}\left[ -2800(c_0 +1)t^3\,+\,140(-1+6 a^2 - 2c_0 +6a^2 c_0 )t^5\right]  \\ 
\\
 \,\,\,\,\,\,\,\,\,\,\,\,\, \,\,\,\,\,\,\,\,\,\,\,\,\, \,\,\,\,\,\,\,\,\,\,\,\,\, + \,\,\,\,
\frac{c_0^2\, a}{16800}\left[+\frac{10}{3}( 66 a^2  - c_0  + 132 a^2 c_0    ) t^7 - \frac{5}{3} \left( 21a^4 - 17 a^2 c_0 + 42 a^4 c_0  \right)t^9  \right] \\
\\
 \,\,\,\,\,\,\,\,\,\,\,\,\, \,\,\,\,\,\,\,\,\,\,\,\,\, \,\,\,\,\,\,\,\,\,\,\,\,\, + \,\,\,\,  \frac{c_0^2\, a}{16800}\left[  - \frac{307}{22}a^4 c_0 t^{11}  + \frac{77}{52} a^6 c_0 t^{13}  \right],
 \end{array}
 \label{Anh11}
\end{equation}

\begin{equation}
\begin{array}{l}
x_4\left( t \right)\,\,=\,\,x_3\left( t \right)\,\,+\,\, \frac{c_0^2\, a}{432432000}\left[ 72072000 (1+c_0)^2t^3 \,-\, 3603600(1+c_0)(-1 6a^2 -3c_0+6a^2c_0)t^5     \right]  \\ 
\\
 \hspace{2.3cm}+ \,\,\,\frac{c_0^2\, a}{432432000}\left[ -85800 (66a^2 -2c_0 264a^2 -3c_0^2 +198 a^2 c_0^2 )t^7 \right]\\
 \\
 \hspace{2.3cm}  +\, \,\frac{c_0^2\, a}{432432000}\left[\frac{3575}{3}(756 a^4 - 1224 a^2 c_0 + 3024 a^4 c_0 + c_0^2 - 1836 a^2 c_0^2 + 2268 a^4 c_0^2      )t^9 \right]\\
 \\
 
 \hspace{2.3cm} +\,\,\frac{c_0^2\, a}{432432000}\left[ 130 a^2 c_0 (5526 a^2 - 461 c_0 + 8289 a^2 c_0) t^{11} \right] \\
 \\
\hspace{2.3cm}-\,\,\frac{c_0^2\, a}{432432000}\left[ \frac{15}{2} a^4 c_0 (10164 a^2 - 9481 c_0 + 15246 a^2 c_0) t^{13} - 
 \frac{40533}{2} a^6 c_0^2 t^{15} + \frac{438669}{272} a^8 c_0^2 t^{17} \right].
 \end{array}
 \label{Anh011}
\end{equation}
\\
Approximation series solution can be obtained to any desired number of terms. We can therefore set up an approximate solution through the following series expansion

\begin{equation}
x\left(t\right)\,\,=\,\, x_0\left(t\right)  \,\, +\,\, x_1\left(t\right)\,\, +\,\, x_2\left(t\right)\,\, +\,\, x_3\left(t\right) \,\, +\,\, x_4\left(t\right) + \,\,...,
 \label{Anh012}
\end{equation} i.e.,

\begin{equation}
\begin{array}{l}
x\left( t \right)\,=\,a t \,+\,\frac{1}{6} \left(4 a c_0 + 6 a c_0^2 + 4 a c_0^3 + a c_0^4\right) t^3\\
\\
\hspace{1.5cm}+\,\, \frac{1}{120} \left(-24 a^3  c_0 + 6 a  c_0^2 - 36 a^3  c_0^2 + 8 a  c_0^3 - 24 a^3  c_0^3 +  3 a  c_0^4 - 6 a^3  c_0^4\right) t^5\\
\\
\hspace{1.5cm}+\,\, \frac{1}{5040}\left(-396 a^3 c_0^2 + 4 a c_0^3 - 528 a^3 c_0^3 + 3 a c_0^4 - 198 a^3 c_0^4\right) t^7 \\ 
\\
\hspace{1.5cm}+\,\, \frac{1}{362880} \left(4536 a^5 c_0^2 - 2448 a^3 c_0^3 + 6048 a^5 c_0^3 + a c_0^4 - 1836 a^3 c_0^4 +  2268 a^5 c_0^4\right)t^9\,+\,\, 0\left[t\right]^{10}.
\end{array}
 \label{Anh12}
\end{equation}

With ${c_0=-1}$, we get

\begin{equation}
x\left( t \right)\,=\,a t \,-\, \frac{a}{6}t^3 \,+\, \frac{1}{120}\left(a + 6 a^3\right) t^5 \,-\,\left(\frac{a}{5040} + \frac{11 a^3}{840}\right) t^7 \,+\, \frac{1}{362880}\left(  a + 612 a^3 + 756 a^5  \right)t^9 \,+\,\,0\left[t\right]^{10} 
 \label{Anh13}.
\end{equation}

Since, the homotopy series solution should fulfills the initial conditions, it is straightforward to realize that value of $a$ has to be $\frac{1}{\sqrt{2}}$. We then get

\begin{equation}
x\left( t \right)\,=\frac{t}{\sqrt{2}}\,-\, \frac{t^3}{6\sqrt{2}}\,+\, \frac{t^5}{30\sqrt{2}} \,-\,\frac{17\,t^7}{2520\sqrt{2}} \,+\, \frac{31\,t^9}{22680\sqrt{2}} \,+\,\, 0\left[t\right]^{10} 
 \label{Anh14},
\end{equation} which can be rewritten as follows

\begin{equation}
x\left( t \right)\,=\,\frac{1}{\sqrt{2}}t\,-\, \frac{1}{3}\left(\frac{1}{\sqrt{2}}t\right)^3\,+\, \frac{2}{15}\left(\frac{1}{\sqrt{2}}t\right)^5 \,-\,\frac{17}{315}\left(\frac{1}{\sqrt{2}}t\right)^7 \, + \, \frac{62}{2835}\left(\frac{1}{\sqrt{2}}t\right)^9\, + \,\, 0\left[t\right]^{10} 
 \label{Anh15}.
\end{equation}
It is then straightforward to conclude that for higher-order approximation

\begin{equation}
x\left( t \right)\,=\,\tanh\left(  \frac{1}{\sqrt{2}} \,t \right)
\label{Anh16}.
\end{equation}

As shown in Table~\ref{tab:anharmonicccpaction}, using our approach based on the optimization of the action, we are able to find the HAM parameter $c_0$ which satisfies Eq.~(\ref{Anh2}), and the convergence to the HAM parameter $c_0$ is fast.  At the  $5^{th}$-order homotopy approximation, $c_0 = -0.9996$, which is already very close to $-1$. As the order of the approximation increases, 
we observe that the HAM parameter converges to $-1$, which is also here the exact value of the parameter $c_0$.

\begin{table}[h!t]
\begin{tabular}{ |p{1.5cm}|p{1.3cm}|p{2.0cm}|p{3.0cm}|}
\hline
\multicolumn{4}{|c|}{HAM parameter $c_0$ at different orders of approximation} \\
 \hline
$m$th-order & \centering{$c_0$}  & \centering{$E\left(t=1, \,\,c_0 \right)$} & $\left| E_{exact} - E_{approximate}\right| $ \\
 \hline
 $1^{st}$   &   -3.7070         & 0.0913             & 0.1586   \\
  \hline
 $2^{nd}$  &   -0.9835        & 0.248721         & 1.27$\times10^{-3}$\\ 
  \hline
 $3^{rd}$   &   -0.9297       & 0.249868    & 1.31$\times10^{-4}$\\
  \hline  
 $4^{th}$   &    -0.9807      & 0.25000526     & $5.26\times10^{-6}$\\
  \hline
 $5^{th}$   &    -0.9996      & 0.249999412       & $5.87\times10^{-7}$\\
  \hline 
\end{tabular}  
\caption{The HAM parameter ${c}_0$ (column 2) with increasing order of the approximation (column 1) for the HAM based on the LAP optimization of the action (with $t=1$). In column 3 we report the energy of the HAM solution. Column 4 shows the error on the energy using our approach. $k$ and $\gamma$ are constants of order unity, and $a=\frac{1}{\sqrt{2}}$.} 
\label{tab:anharmonicccpaction}
\end{table}

\newpage


\begin{figure}[ht]
\centering
\includegraphics[width=2.3in,angle=0]{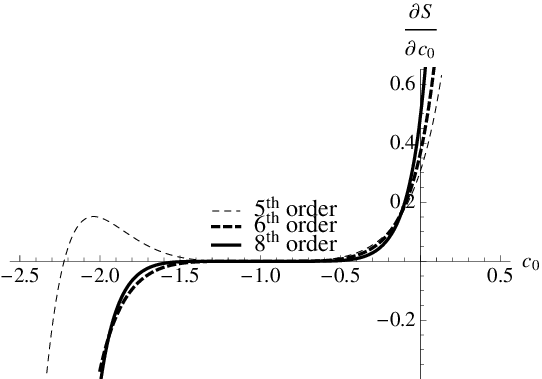}(a)\,\,\,\,\,\,\includegraphics[width=2.3in,angle=0]{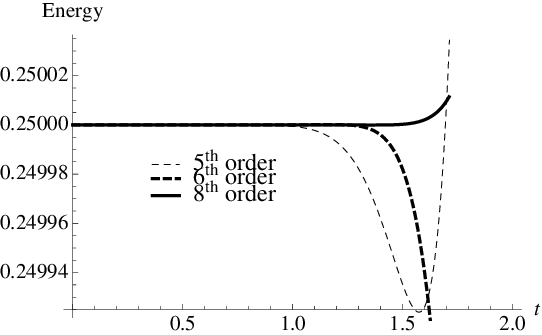}(b)
\includegraphics[width=2.3in,angle=0]{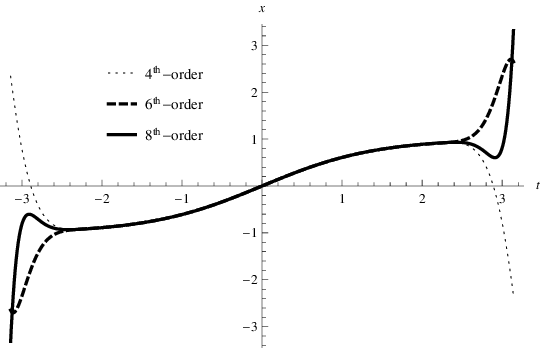}(c)\,\,\,\,\,\,\includegraphics[width=2.3in,angle=0]{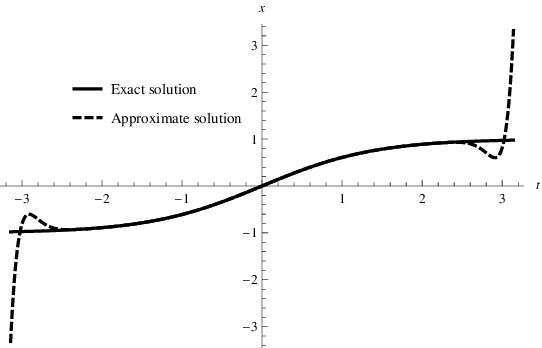}(d)
\caption{Panel (a) shows the derivative the action with respect to $c_0$, $\frac{\partial\,S}{\partial\,c_0}$, at different orders of HAM approximation with LAP with $t=1$. Panel (b) reports the energy vs. time showing 
energy conservation up to a time $t$ which increases with the orders of HAM approximation. Panel (c) shows the HAM approximation of the solution at increasing order. In panel (d) we compare the exact solution with the $8^{th}$-order of approximation. $k$ and $\gamma$ are constants of order unity, and $a=\frac{1}{\sqrt{2}}$.} 
\label{fig:quarticsolutions}
\end{figure}

Using our approach, we are able to find the HAM parameter, $c_0$, that allows for the best energy conservation of the quartic oscillator up to a certain final time, and consequently, we got very close to the exact solution $x\left(t\right)=\tanh\left(\frac{1}{\sqrt{2}}t\right)$.  We also see that the flatness (giving the the set of HAM parameters $c_0$ extremizing the approximated action) increases with the order of the approximation (see Fig.~\ref{fig:quarticsolutions}-a). 
\\

Let us move forward by investigating a case of nonlinear problem, where the 
solution is 
not simply expressed in terms of simple analytical functions. In this case the HAM parameter, $c_0$, that governs the solution cannot be easily guessed. 

\subsection{Cubic anharmonic oscillator.}
\label{cubicOscillatorySysytem} 

\begin{figure}[ht]
\centering
\includegraphics[width=3.in,angle=0]{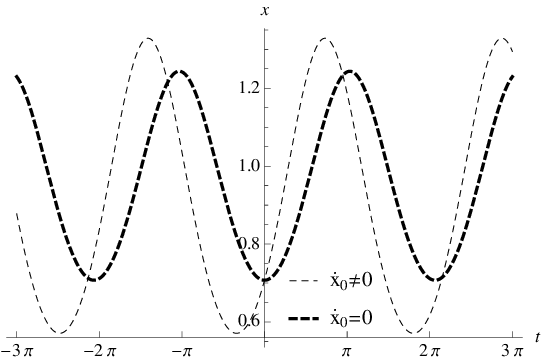}
\caption{Numerical solution of the cubic anharmonic oscillator problem. 
The thick dashed plot represents the exact solution with zero velocity $(\dot{x}_0=0)$, while the dashed plot gives the solution with nonzero velocity $(\dot{x}_0=\frac{1}{4})$.} 
\label{fig:numericalcubic}
\end{figure}

We now consider 
a one-dimensional cubic anharmonic oscillator system. 
Its Hamiltonian reads

\begin{equation}
 H\left(x, \, p \right)=\frac{p^2}{2} \,\, -  \,\, \frac{1}{2}\,x^2  \,\, + \,\, \frac{\gamma}{3}\,x^3
 \label{cubic1},
\end{equation} 
with the mass ${\cal M}$ of the system being again taken as one. For $\gamma>0$ the dynamics of the system is stable in the region $0<x_{0}<\frac{3}{2\gamma}$ for $\dot{x}_{0}=0$, and for $\dot{x}_{0}\neq0$, $x_{0}$ and $\dot{x}_{0}$ are bounded by the curve described by the following equation: $\dot{x}_{0}=\pm x_{0}\sqrt{1 - \frac{2\gamma}{3}x_{0}}$. 
With $\gamma=1$ the equation of motion reads

\begin{equation}
x_{tt} \,\,-\,\,x \,\,+\,\,x^2 =0
 \label{cubic2}.
\end{equation}
The numerical solution of Eq.~(\ref{cubic2}) is given in Fig.~\ref{fig:numericalcubic}.


Proceeding as before, we get for $m=1,\cdots,5$ 

\begin{equation}
x_0\,\,=\,\,\frac{1}{\sqrt 2}  
 \label{cubic3},
\end{equation}

\begin{equation}
x_1\,\,=\,\,x_0 \,\,+\,\,\frac{1}{4}\left( -c_0 + \sqrt{2} \,c_0\right)t^2
 \label{cubic4},
\end{equation}

\begin{equation}
x_2\,\,=\,\,x_1 \,\,+\,\,\frac{1}{4}\left(c_0^2 - \sqrt{2}\,c_0^2\right)\,t^2 \,\,+\,\,\left(-\frac{c_0^2}{16} \,+\,\frac{c_0^2}{12\sqrt{2}}\right)t^4
 \label{cubic5},
\end{equation}

\begin{equation}
x_3\,\,=\,\,x_2 \,\,+\,\,\frac{1}{4}\left(c_0^3 - \sqrt{2}\,c_0^3\right)\,t^2 \,\,+\,\,\left(-\frac{c_0^3}{8} \,+\,\frac{c_0^3}{6\sqrt{2}}\right)t^4 \,\,+\,\,\left(\frac{c_0^3}{90} \,-\,\frac{11 c_0^3}{720\sqrt{2}}\right)t^6
 \label{cubic6},
\end{equation}

\begin{equation}
x_4\,\,=\,\,x_3 \,\,+\,\,\frac{1}{4}\left(c_0^4 - \sqrt{2}\,c_0^4\right)\,t^2 \,\,+\,\,\left(-\frac{3 c_0^4}{16} \,+\,\frac{c_0^4}{4\sqrt{2}}\right)t^4 \,\,+\,\,\left(\frac{c_0^4}{30} \,-\,\frac{11 c_0^4}{240\sqrt{2}}\right)t^6\,\,+\,\,\left(\frac{-143c_0^4}{80640} \,+\,\frac{ 17 c_0^4}{6720\sqrt{2}}\right)t^8
 \label{cubic66},
\end{equation}

\begin{equation}
\begin{array}{l}
x_5\,\,=\,\,x_4 \,\,+\,\,\frac{1}{4}\left(c_0^5 - \sqrt{2}\,c_0^5\right)\,t^2 \,\,+\,\,\left(-\frac{19 c_0^5}{96} \,+\,\frac{13 c_0^5}{48\sqrt{2}}\right)t^4 \,\,+\,\,\left(\frac{71 c_0^5}{960} \,-\,\frac{17 c_0^5}{160\sqrt{2}}\right)t^6\,\,+\,\,\left(\frac{-2629c_0^5}{161280} \,+\,\frac{619 c_0^5}{26880\sqrt{2}}\right)t^8\\
\\
\hspace{1.5cm}+\,\,\,\,\,\, \left(\frac{-17909 c_0^5}{14515200}\,-\,\frac{4213 c_0^5}{2419200\sqrt{2}}\right)t^{10} .
 \end{array}
\label{cubic666}
\end{equation}
We can then set up an approximate solution through the following series expansion

\begin{equation}
x\,\,=\,\,x_0\,\, +\,\,x_1+\,\,x_2\,\, +\,\,x_3\,\, +\,\,x_4\,\, +\,\,x_5 +\,\,...
 \label{Appppp}
\end{equation}

\begin{figure}[ht]
\centering
\includegraphics[width=2.3in,angle=0]{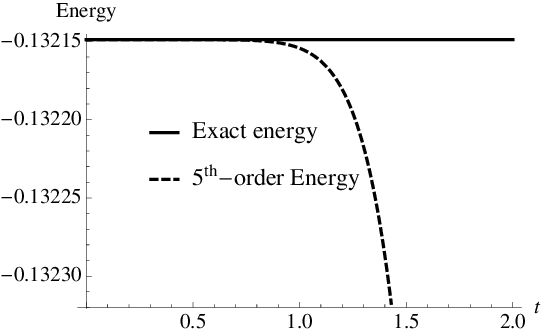}(a)\,\,\,\,\,\,\includegraphics[width=2.3in,angle=0]{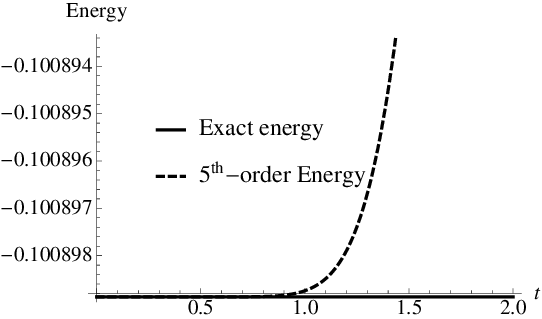}(b)
\includegraphics[width=2.3in,angle=0]{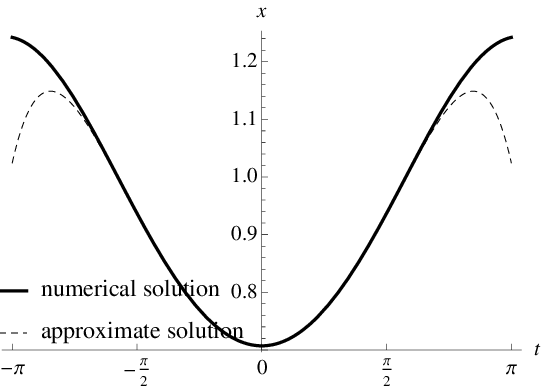}(c)\,\,\,\,\,\,\includegraphics[width=2.3in,angle=0]{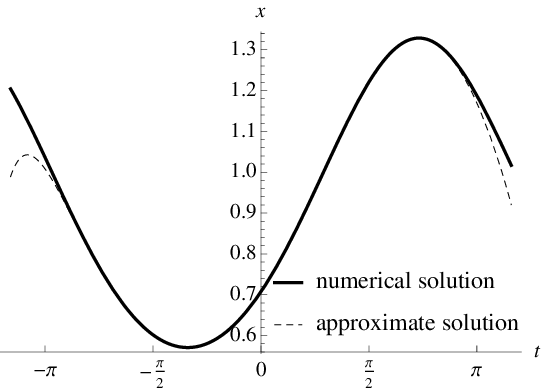}(d)
\caption{Panels (a) and (b) show the conservation of the 
energy of the system at a finite time with zero velocity and nonzero velocity, respectively.  In panels (c) and (d) we compare the 
numerical solution (Fig.~\ref{fig:numericalcubic}) to the $5^{th}$-order of approximate solutions, with zero velocity and nonzero velocity, respectively.} 
\label{fig:Zerocubicsolution}
\end{figure}

As shown in Tables~\ref{tab:ccubicaction} and~\ref{tab:ccubicactionn}, our approach permits to find the HAM parameter $c_0$ at different orders of approximation both for zero and nonzero initial velocites, respectively. We find that $c_0$ and the energy at time $t=1$ stabilize around a value. We see that in this case it is not necessary to go to a high order of approximation to get rather close to the exact solution. At  $5^{th}$ order of approximation, we approach the exact solution (see Fig.~\ref{fig:Zerocubicsolution}).

\begin{table}[h!t]
\begin{tabular}{ |p{1.5cm}|p{1.2cm}|p{2.0cm}|p{2.9cm}|}
\hline
\multicolumn{4}{|c|}{Cubic oscillator system with initial zero velocity} \\
 \hline
$m$th-order & \centering{$c_0$}  & \centering{$E\left(t=1, \,\,c_0 \right)$} & $\left| E_{exact} - E_{approximate}\right| $ \\
 \hline
 $1^{st}$   &   0.5395           &-0.113747             & 0.018400 \\
  \hline
 $2^{nd}$  &   -0.9668         &-0.1317467           & 0.0004020  \\ 
  \hline
 $3^{rd}$   &   -0.9104        & -0.1320634     & 0.0000853\\
  \hline    
 $4^{th}$   &   -0.9856        & -0.132150450         & $1.580\times10^{-6}$  \\
  \hline
 $5^{th}$   &   -0.9917        & -0.132154362      & $5.492\times10^{-6}$  \\
  \hline 
\end{tabular}  
\caption{The HAM parameter ${c}_0$ (column 2) with increasing order of the approximation (column 1) for the HAM based on the optimization of the action at time $t=1$. In column 3 we report the energy of the HAM solution at time $t=1$. Column 4 shows the variation of the energy using our approach. Here, the initial conditions are the following: $x_{0} =\frac{1}{\sqrt 2}$ and  $\dot{x}_0=0$, i.e, initial zero velocity. }
\label{tab:ccubicaction}
\end{table}

\begin{table}[h!t]
\begin{tabular}{ |p{1.5cm}|p{1.2cm}|p{2.0cm}|p{2.9cm}|}
\hline
\multicolumn{4}{|c|}{Cubic oscillator system with initial  nonzero velocity} \\
 \hline
$m$th-order & \centering{$c_0$}  & \centering{$E\left(t=1, \,\,c_0 \right)$} & $\left| E_{exact} - E_{approximate}\right| $ \\
 \hline
 $1^{st}$   &    3.1710           & -0.11508     & 0.01418 \\
  \hline
 $2^{nd}$  &   -0.9394         & -0.09992      & $9.78\times10^{-5}$\\ 
  \hline
 $3^{rd}$   &    -1.0080        & -0.1009544    & $5.56\times10^{-5}$ \\
  \hline    
 $4^{th}$   &    -0.9779        & -0.10090154      & $2.67\times10^{-6}$ \\
  \hline
 $5^{th}$   &    -1.0001        & -0.10089872     & $1.39\times10^{-7}$ \\
  \hline 
\end{tabular}  
\caption{
The same as in Table \ref{tab:ccubicaction} but now with $\dot{x}_0=\frac{1}{4}$, i.e, initial nonzero velocity.}
\label{tab:ccubicactionn}
\end{table}

Fig.~\ref{fig:Zerocubicsolution} shows that HAM approximate solution obtained using the LAP is already accurate at the $5^{th}$-order approximation, in its first period. Also, the 
energy of the system is well conserved 
at the considered 
final time. When such time increases, then one has to increase the order of the approximation. 


\subsection{The Korteweg-de Vries (KdV) equation}
\label{KdV equation}

We have computed numerically the action of approximate solutions of linear and nonlinear Hamiltonian ODE and found that the action is rather flat in the parameter that controls the approximation. 
Now, we are interested in extending the method to Hamiltonian PDE. For this purpose, we consider 
the well-known KdV equation \cite{Miles2006, Matteo2021} using the LAP.

Let us consider the one-dimensional KdV equation 

\begin{equation}
u_t - 6 u u_x + u_{xxx} = 0
\label{kdv1},
\end{equation} 
where  $u(x,\,t)$ is the wave function. The traveling wave soliton of the KdV equation is given by 

\begin{equation}
u(x,\,t)=-\frac{v}{2}sech^2\left[ \frac{\sqrt{v}}{2} \left( x-v t -x_0\right) \right]
\label{kdv2},
\end{equation} 
where $v$ and $x_0$ represent the wave velocity and the integration constant, respectively.

To construct the Lagrangian density, we introduce the Lagrangian (density) of KdV equation $L=L(u, u_t, u_x, u_{xx}; t,x)$ where $u(x,\,t)=\phi_{x}(x,\,t)$. Minimising the action $S=\int{L\,dt}$ yields

\begin{equation}
\frac{\partial L}{\partial \phi} - \frac{\partial }{\partial t}\left(\frac{\partial L}{\partial \phi_t}\right) - \frac{\partial }{\partial x}\left(\frac{\partial L}{\partial \phi_x}\right) + \frac{\partial^2 }{\partial x^2}\left(\frac{\partial L}{\partial \phi_{xx}}\right)=0.
\label{kdv3}
\end{equation}
The Lagrangian giving Eq.~(\ref{kdv1}) is  

\begin{equation}
L=-\frac{1}{2}\phi_x \phi_t - \phi_{x}^3 - \frac{1}{2}\phi_{xx}^2
\label{kdv4}.
\end{equation}
Using Eq.~(\ref{kdv4}) to derive the Hamiltonian density, by which the energy of KdV equation is calculated, one gets

\begin{equation}
H= \Pi\phi_t + 2\frac{\partial}{\partial t}\left( \frac{\partial L}{\partial \phi_{xx}}\right)\phi_t - L
\label{kdv5},
\end{equation} 
where the density of its conjugate momentum is $\Pi=\partial L/\partial \phi_t$.  The reader is referred to Ref.~\cite{Matteo2022, Gelfand2000} for details on the Hamiltonian field theory close to the wave
equation, and to \cite{Mackay1983} for dynamical stability in Lagrangian and Hamiltonian systems. Substituting Eq.~(\ref{kdv4}) into Eq.~(\ref{kdv5}) leads to

\begin{equation}
H= - \phi_{xxt}\phi_{t} + \phi_{x}^3 + \frac{1}{2}\phi_{xx}^2
\label{kdv6}.
\end{equation}
Then, the energy of static soliton in KdV equation is calculated by using the expression

\begin{equation}
E = \int_{-\infty}^{\infty} {H(x,0)} \,dx= \int_{-\infty}^{\infty} {\left( \phi_{x}^3 + \frac{1}{2}\phi_{xx}^2\right)} \,dx 
\label{kdv7},
\end{equation} where we set $t=0$ in  Eq.~(\ref{kdv2}) into  Eq.~(\ref{kdv7}). For $v=1$ and $x_0=0$ as an example we 
find 
$E=-0.2$. 
Note that the initial energy is time-independent since $\phi_t$ in Eq.~(\ref{kdv6}) always gives zero.

Let us now find the solution of Eq.~(\ref{kdv1}) by means of the HAM. For this purpose, we first perform the following Galilean transformation:

\begin{equation}
u\left(x,\,t\right)\,=\,u\left( \eta \right)\,\,\,\,\,\,\,\,\,\,\,\,\,\,\,\,\,\,\,\,\,\,\, \textrm{where} \,\,\,\,\,\,\,\,\,\,\,\,\, \eta=x - v\,t
 \label{kdv9},
\end{equation} 
with $v$ the group velocity of the wave. Therefore Eq.~(\ref{kdv1}) can be rewritten as follows: 

\begin{equation}
v u_{\eta} \,-\, 6 u u_{\eta} \,-\,u_{\eta\eta\eta} =0 
 \label{kdv10}.
\end{equation}
Performing an integration and setting 
the integration constant to zero, we obtain

\begin{equation}
v u \,-\, 3 u^2 \,-\,u_{\eta\eta} =0 
 \label{kdvv}.
\end{equation}
In order to perform the HAM scheme, we choose the following initial guess and auxiliary function:

\begin{equation}
u_0\left( \eta \right)=-\frac{v}{2}, \,\,\,\,\,and \,\,\,\,\,  \mathcal{H}\left( t \right)=1,
 \label{kdv11}
\end{equation}
respectively. Since the linear operator $\mathcal{L}$ should be chosen in order to permit the initial approximation, we set 

\begin{equation}
\mathcal{L}\left[  u \right]\,\equiv \, u_{\eta\eta}.
 \label{kdv12}
\end{equation} 
The solution can be written in the following way 

\begin{equation}
u\left(\eta \right)= u_0\left(\eta  \right)\,\,+\,\,\sum\limits_{m=  1}^{M } { u_m\left(\eta  \right) }, 
 \label{KDV}
\end{equation} 
where

\begin{equation}
u_m\left(\eta  \right) \,\,=\,\chi_m\,\,u_{m-1}\left(\eta  \right) \,\,+\,\, c_0 \, \mathcal{L}^{-1}\left[  \mathcal{R}\left( u_{m-1}\left(\eta  \right)   \right)   \right]
 \label{kdv13},
\end{equation} and 

\begin{equation}
\mathcal{R}\left( u_{m-1} \left(\eta  \right)   \right) \,\,=\,u_{({m-1})\eta\eta} \,\,-\,\, v \, u_{m-1}\,\,-\,\,3\sum\limits_{i=  0}^{m-1 } { u_i\,u_{m-1-i} }
 \label{kdv14}.
\end{equation}

Combining Eqs.~(\ref{kdv11}),~(\ref{kdv12}),~(\ref{KDV}),~(\ref{kdv13}), and~(\ref{kdv14}), one gets for $m=1,\cdots,5$

\begin{equation}
u_1\left( \eta \right)= - \,c_0\frac{\eta^2 }{8}
 \label{approx1},
\end{equation}

\begin{equation}
u_2\left( \eta \right)= u_1\left( \eta \right) \,-\, \frac{1}{12}c_0^2\left( \frac{3 \eta^2}{2} \,+\, \frac{\eta^4}{4}\right)
 \label{approx2},
\end{equation}

\begin{equation}
u_3\left( \eta \right)=u_2\left( \eta \right) \,-\, \frac{1}{960}c_0^2\left( 120(1+c_0)\eta^2 + 20(1+2c_0)\eta^4 + \frac{17c_0 \eta^6}{6}\right)
 \label{approx3},
\end{equation}

\begin{equation}
u_4\left( \eta \right)=u_3\left( \eta \right) - \frac{c_0^2\left(  2520(1+c0)^2\eta^2 + 420(1+4 c_0 + 3 c_0^2)\eta^2 +\frac{119}{2}c_0 (2+3c_0)\eta^6 + \frac{31c_0^2 \eta^8}{4} \right)}{20160}
 \label{approx4},
\end{equation}

\begin{equation}
\begin{array}{l}
u_5\left( \eta \right)=u_4\left( \eta \right) \,- \,\frac{181440(1+c_0)^3\eta^2 + 30240(1+c_0)^2(1+4c_0)\eta^4 +12852c_0(1+3c_0 +2 c_0^2)\eta^6 + 558c_0^2(3+4c_0)\eta^8 + \frac{691 c_0^3 \eta^{10}}{10}}{1451520}
 \label{approx5}.
\end{array}
\end{equation}
The approximation series solution can be obtained to any desired number of terms, we can therefore set up an approximate solution through the following series expansion

\begin{center}
$u\left( \eta \right)\,\,=\,\, u_0\left( \eta \right) \,\, +\,\, u_1\left( \eta \right)\,\, +\,\, u_2\left( \eta \right)\,\, +\,\, u_3\left( \eta \right) \,\,+  u_4\left( \eta \right)\,\,+\,\,u_5\left( \eta \right) + ...$
\end{center}

The $12^{th}$-order approximate homotopy solution reads

\begin{equation}
\begin{array}{l}
u\left( \eta \right)=-\frac{v}{2} + \frac{v^2\eta^2}{8} - \frac{v^3\eta^4}{48} + \frac{17 v^4\eta^6}{5760} - \frac{31v^5\eta^8}{80640} + \frac{691 v^6 \eta^{10}}{14515200 } - \frac{5461 v^7 \eta^12}{958003200} + \frac{929569 v^8 \eta^{14}}{1394852659200} \\
\\
 \hspace{1.1cm} - \frac{3202291 v^9 \eta^{16}}{41845579776000} + \frac{221930581 v^{10}\eta^{18}}{25609494822912000} - \frac{4722116521v^{11}\eta^{20}}{4865804016353280000} + \frac{56963745931 v^{12}\eta^{22}}{528941518954168320000} \\
\\
 \hspace{1.1cm} - \frac{14717667114151 v^{13}\eta^{24}}{1240896803466478878720000} 
\end{array}
 \label{approx6},
\end{equation} where we have assumed ${c_0=-1}$. It is straightforward to conclude that for higher-order approximation we obtain

\begin{equation}
u\left( \eta \right)\,=-\frac{v}{2}\,sech^2\left( \frac{\sqrt{v}}{2}\,\eta \right),
 \label{approx7}
\end{equation} 
i.e.,

\begin{equation}
u\left(x\, ,t\right)\,=-\frac{v}{2}\,sech^2\left( \frac{\sqrt{v}}{2}\left(x - v\,t\right)\right)
 \label{approx8}.
\end{equation}

Table~\ref{static} gives at each order of the homotopy approximation the optimal HAM parameter, the corresponding energy of static soliton and the absolute error between the exact energy and the approximate one:

\begin{equation}
\Delta E= \mid  E_0\left( \phi_{exact}\right) -  E_0\left( \phi_{appr}\right) \mid 
\label{kdvvvv}.
\end{equation} 
Here

\begin{equation}
 E_0\left( \phi_{appr}\right) =  E\left( \phi_{appr}\left(x,t=0\right)\right) =\int_{-2\pi}^{2\pi} {\left[ \left(\frac{\partial \phi_{appr}}{\partial x}\right)^3 + \frac{1}{2}\left(\frac{\partial^2 \phi_{appr}}{\partial  x^2}\right)^2\right]_{t=0}} \,dx 
\label{kdvvv7}.
\end{equation}


\begin{table}[h!t]
\begin{tabular}{ |p{1.5cm}|p{1.3cm}|p{2.3cm}|p{2.7cm}| }
 \hline
$m$th-order  &\centering{$c_0$} & energy {$E_0$} of the static solution  & absolute error $\Delta E$, see 
(\ref{kdvvvv})\\
  \hline
 $2^{nd}$  &     -0.249           &         -0.126  &      0.073      \\
  \hline  
 $3^{rd}$   &    -0.1788           &     -0.2808       &      0.0808       \\
  \hline 
 $4^{th}$   &   -0.2052            &    -0.24205       &        0.04205   \\
  \hline  
 $5^{th}$   &    -0.2072           &     -0.2277        &        0.02770    \\ 
  \hline
 $6^{th}$   &     -0.2300          &  -0.21385          &       0.01385      \\
 \hline
 $7^{th}$   &     -0.2321         &     -0.20897      &       0.00897      \\
 \hline
 $8^{th}$  &    -0.2452           &   -0.20485         &  0.00485 \\
 \hline
 $9^{th}$   &       -0.2467        &   -0.20315       &  0.00315 \\
 \hline
 $10^{th}$   &    -0.2550           &  -0.2017          &    0.0017    \\
 \hline
 $11^{th}$  &     -0.2610        &    -0.2003     &        0.0003     \\
 \hline
\end{tabular}
\caption{The HAM parameter $c_0$ (column 2) at increasing order of the approximation (column 1) with $t=1$ (and $v=1$). The energy of the static soliton solution of the KdV equation and the absolute error $\Delta E$ are reported in columns 3 and 4. One sees that 
the HAM parameter $c_0$ does not quickly converge to its optimal value $c_0=-1$, but at variance the energy of the static soliton solution converges quickly to its exact value $ E_0 = -0,2$. }
\label{static}
\end{table}

\begin{figure}[ht]
\centering
\includegraphics[width=2.5in,angle=0]{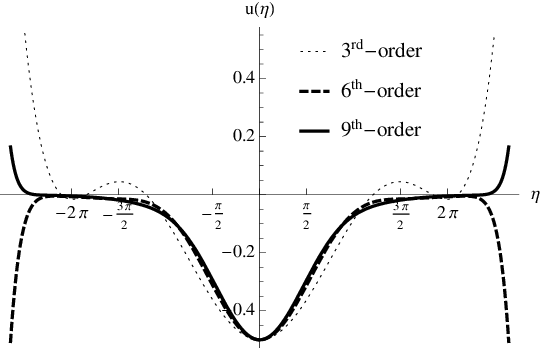}(a)\,\,\,\,\,\,\includegraphics[width=2.5in,angle=0]{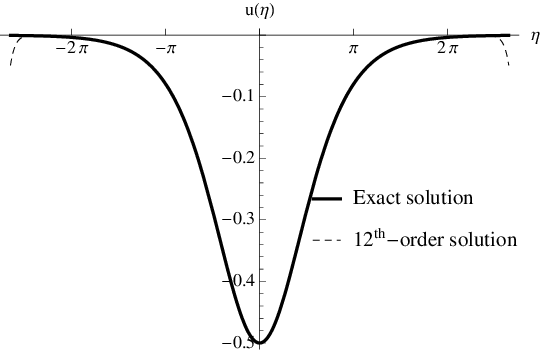}(b)
\caption{Static soliton solution for KdV equation plotted within the period of the soliton wave. Panel (a) shows the convergence of the numerical solutions when the order of the approximation is increased. Panel (b) shows the comparison of the exact solution with the $12^{th}$-order HAM solution using LAP. The group velocity of the wave, $v$, is taken as one.}
\label{fig:statickdvsoliton}
\end{figure}

Fig.~\ref{fig:statickdvsoliton} compares the static solution of the KdV equation and the solution obtained combining the LAP to the HAM. 

We are now in position to compare the LHAM method with the residual error square method discussed in section (\ref{homotopydescription}). The result of our comparison is presented in Table~\ref{actionVSresidual}, where we refer 
to the residual error square method as the ``Standard-HAM". We remark that the HAM to calculate the approximate solution for $u(x,t)$ is the same, i.e., with the same auxiliary linear operator $\mathcal{L}$ and the same initial guess of the solution. One could improve the outcome of the HAM calculation (and obtaining values of $c_0$ closer to the exact one $c_0=-1$) by improving such a choice. However, here the goal is to show an example of comparison of determining the optical $c_0$ in the two different ways, the difference being in the way of fixing $c_0$ my minimizing the residual error square (Standard-HAM) or using the LAP (LHAM). We see that we have better result using the approach based on the LAP, and we also mention that using standard mathematical software we find that is more difficult for larger order of approximation to determine the optimal $c_0$ by the Standard-HAM (we were unable to find the optimal $c_0$ for the $11^{th}$ order of approximation using the Standard-HAM). We refer to the Appendices for the comparison of LHAM and Standard-HAM for the harmonic and anharmonic oscillators.

\begin{table}[h!t]
\begin{tabular}{ |p{2.5cm}|p{1.5cm}|p{3.2cm}|p{3.5cm}| }
 \hline
         & \centering{$c_0$ } & energy of the static solution  \centering{$E_0$}  & absolute error $\Delta E_0 $, see 
         (\ref{kdvvvv})\\
 \hline
 {\bf $5^{th}$-order}\\
 
  \hline
LHAM            & -0.2072        &  -0.2277      &     0.02770    \\
\hline
Standard-HAM                                & -0.1820       &  -0.2376          &   0.03767 \\
 \hline
 
   {\bf  $7^{th}$-order  }                                              \\
 \hline
LHAM            &   -0.2321        &  -0.20897   & 0.00897    \\
\hline
Standard-HAM                                & -0.1977         &  -0.21515    & 0.01515  \\
  \hline 
   {\bf  $11^{th}$-order  }            \\
 \hline
LHAM           &    -0.2610       &  -0.2003   &  0.0003   \\
\hline
Standard-HAM                               & ---                       &  ---               & ---     \\
  \hline 
\end{tabular}  
\caption{The HAM parameter ${c}_0$ (column 2) with increasing order of the approximation (column 1) for the HAM method based on the optimization of the action and for the standard HAM based on the residual error square minimization. In column 3 we report the energy of the static soliton solution. Column 5 shows the absolute error $\Delta E_0 $ for our approach and for the standard HAM using the minimisation of the residual error square. The group velocity of the wave, $v$, is taken as one.}
\label{actionVSresidual}
\end{table}


By the analysis of the comparison between LHAM and the residual error square method, 
we conclude that implementing the extremization of the action and the optimal conservation of the energy, our method accelerates the convergence (and reduce the time-consume) of the approximate solution. 



\section{Concluding remarks and perspectives}
\label{conclusions}

The Homotopy Analysis Method (HAM)~\citep{SJLiao2003,SJLiao2014} has been widely used to obtain approximate numerical solutions of nonlinear problems. In this paper, we have combined the Least Action Principle (LAP) with the HAM in order to find a better estimation of the HAM parameter $c_0$ for systems for which one can define the Lagrangian and the Hamiltonian. In our method, to which we refer to as LHAM, the LAP allows us to obtain several values of the HAM parameter $c_0$ which extremize the action. Among these values, we choose 
the one that better conserve the energy. 
This simple method accelerates the convergence of the approximate solution to the exact one 
and the Korteweg-de Vries (KdV) partial differential equation.  Our method is found to be efficient also when non-linearity is strong. 

Since the Lagrangian and Hamiltonian dynamics are applicable  to many different and relevant natural systems, 
our improved method can be useful in many contexts. As first, as future work one should systematically compare with other approaches implementing energy conservation \cite{Sanz-serna1994}. Moreover, among the many example of Hamiltonian dynamics which one could investigate with our approach, let us quote for example 
(i)~the positioning of geosynchronous and GPS satellites \cite{Alessandra2014}, (ii)~ the characterization of Rydberg atoms \cite{Gallagher1994}, and (iii)~the study of multiferroic materials \cite{Spladin2005, Eerenstein2006, Chotorlishvili2013, Paglan2019}.

Our approach can be extended to fractional ordinary and partial differential equations that can be derived from an action principle. 
However, the crucial issue of evaluating the action for these equations deserves further investigations. An interesting class of fractional differential equations describes the continuum limit of the generalized Fermi-Pasta-Ulam-Tsingou (FPUT) lattices with long-range interactions \cite{chendjou2018}. The numerical solutions obtained using our method could be then compared with exact solutions 
\cite{chendjou2019-2}. 


\section*{Acknowledgements}
Thanks to Matteo Gallone and Yu Zhou for valuable discussions and careful reading of the paper. One of authors, G N B Chendjou, acknowledges the hospitality of the Abdus Salam International Centre for Theoretical Physics (ICTP) and the International School for Advanced Studies (SISSA) in Trieste, Italy, where part of the work reported here was carried out.


\appendix
\label{appendix}

\section{Harmonic oscillator}

In this appendix and the following ones, the residual error square corresponding to any given HAM approximate solution ($x_{approximate}\left(t, c_0\right)$) is written as follows: 

\begin{equation}
\epsilon\left(c_0\right)\,=\, \int_{0}^{t}\left({\mathcal{N}\left[ { x_{approximate}\left(t, c_0\right)}\right]}\right)^2 d t 
 \label{appendixharm},
\end{equation} 
where $\mathcal{N}$ is the nonlinear operator related to the equation of motion of a 
problem. 
The residual error square using LHAM corresponds to $\epsilon\left(c_0\right)$ with $c_0$ obtained using Eq.~(\ref{homotopic3}), while the standard residual error square corresponds to $\epsilon\left(c_0\right)$ with $c_0$ obtained using Eq.~(\ref{HAM14}).

We first compare our HAM method based on the optimization of the action with the standard HAM based on the residual error square minimization for the harmonic oscillator. 

\begin{table}[h!t]
\begin{tabular}{ |p{0.8cm}|p{2.3cm}|p{3.2cm}|p{2.5cm}|p{3.cm}|p{2.4cm}|  }
\hline
 & \centering{${{c}_0}$ using LHAM, see (\ref{homotopic3})} &  \centering{$\left| E_{exact} - E\left(t=1, \,\,c_0 \right)\right| $} &\centering{${{c}_0}$ using the standard-HAM, see (\ref{HAM13})}  &$\left| E_{exact} - E\left(t=1, \,\,c_0 \right)\right| $ \\
 \hline
 
  \hline
 $1^{st}$      & 0.5945       &0.5181                              & -0.8433                 & 0.02282 \\
  \hline
 $2^{nd}$    & -0.9477      &$4.3\times10^{-3}$           & -0.9691                 & $5.0\times10^{-4}$ \\
 \hline
$3^{rd}$      & -0.9804      & $8.0\times10^{-5}$         & -0.9891                  &$1.0\times10^{-5}$\\
 \hline
$4^{th}$      & -0.9925      & 5.08$\times10^{-7}$      & -0.9947                   &1.08$\times10^{-7}$ \\
\hline
$5^{th}$      & -0.9959     & $2.69\times10^{-9}$       & -0.9949                   &3.18$\times10^{-9}$ \\
\hline
$6^{th}$      & -0.9977     & 8.91$\times10^{-12}$    & -0.9953                   &6.79$\times10^{-12}$ \\
 \hline
\end{tabular}
\caption{The HAM parameter ${c}_0$ (column 2) with increasing order of the approximation (column 1) with $t=1$ (and $\omega=1$). In column 3 we report the variation of the energy at time $t=1$. Columns 4 and 5 show the HAM parameter $c_0$ and the variation of the energy using the residual error square method with $t=1$ (and $\omega=1$).} 
\end{table}

\begin{table}[h!t]
\begin{tabular}{|p{0.8cm}|p{2.3cm}|p{3.0cm}| p{2.8cm}|p{2.8cm}|}
\hline
& {\centering{Residual error square using LHAM, see (\ref{homotopic3}) and (\ref{appendixharm}) }}  & {\centering{ Standard residual error square, see (\ref{HAM13}) and (\ref{appendixharm})}} & { \centering{$E\left(t=1, \,\,c_0 \right)$ using LHAM, see (\ref{homotopic3})}} & { \centering{$E\left(t=1, \,\,c_0 \right)$ using the standard residual error square, see (\ref{HAM13})}} \\ 
\hline
 $1^{st}$   &   2.8760                                            & 0.0160                              &    1.0181                               &   0.5228 \\
 \hline
  $2^{nd}$ &  0.000075                                    &  0.000011                      &     0.50433                                    &  0.5005 \\
 \hline
 $3^{rd}$   &   3.7$\times10^{-8}$                      &  2.4$\times10^{-9}$           &   0.499919                              &  0.49998    \\
 \hline
  $4^{th}$  &   1.71$\times10^{-12}$                     &  1.70$\times10^{-13}$    &   0.5000005                            &  0.50000010       \\
 \hline
  $5^{th}$  &   1.13$\times10^{-13}$                       &  2.27$\times10^{-13}$    &    0.499999997                 & 0.499999996  \\
 \hline
\end{tabular}
\caption{Residual error square for increasing order of the approximation using the least action principle (column 2) or the residual error square method (column 3) with $t=1$ (and $\omega=1$). The corresponding energies at $t=1$ are reported in columns 4 and 5. }
\end{table}

\section{Quartic oscillator}
\label{Quarticoscillator}

Here, we compare our HAM method based on the optimization of the action with
the standard HAM based on the residual error square minimization for the quartic anharmonic oscillator. 

\begin{table}[h!t]
\begin{tabular}{ |p{.8cm}|p{2.3cm}|p{3.2cm}|p{2.5cm}|p{3.0cm}|p{2.4cm}|  }
\hline
 & \centering{${c_0}$ using LHAM, see (\ref{homotopic3})} &  \centering{$\left| E_{exact} - E\left(t=1, \,\,c_0 \right)\right| $} &${c_0}$ using the standard-HAM, see (\ref{HAM13}) &$\left| E_{exact} - E\left(t=1, \,\,c_0 \right)\right| $ \\
 \hline
 
  \hline
 $1^{st}$      & -3.7070          &   0.1586                           &  -0.9884                & 2.7$\times10^{-4}$\\
  \hline
 $2^{nd}$    & -0.9835        &   1.2$\times10^{-3}$          &  -0.8861                 &  5.6$\times10^{-4}$\\
 \hline
$3^{rd}$      & -0.9297         &  1.3$\times10^{-4}$         & -0.9740                   & 5.5$\times10^{-6}$\\
 \hline
$4^{th}$      &  - 0.9807      & 5.2$\times10^{-6}$           &  -1.003                    & 3.5$\times10^{-7}$      \\
\hline
$5^{th}$      &  -0.9996      &  5.8$\times10^{-7}$           &  -0.9727                    & 2.7$\times10^{-8}$\\
\hline
\end{tabular}
\caption{The HAM parameter $c_0$ (column 2) with increasing order of the approximation (column 1). In column 3 we report the variation of the energy at time $t=1$ with $a=\frac{1}{\sqrt{2}}$. Columns 4 and 5 show the HAM parameter $c_0$ and the variation of the
energy using the residual error square method. $k$ and $\gamma$ are constants of order unity.} 
\end{table}

\begin{table}[h!t]
\begin{tabular}{|p{0.8cm}|p{2.3cm}|p{3.0cm}|p{2.8cm}|p{2.8cm}|}
\hline
& {\centering{Residual error square using LHAM, see (\ref{homotopic3}) and (\ref{appendixharm})} }  & { \centering{Standard residual error square, see (\ref{HAM13}) and (\ref{appendixharm})}} & { \centering{$E\left(t=1, \,\,c_0 \right)$ using LHAM, see (\ref{homotopic3})}} & { \centering{$E\left(t=1, \,\,c_0 \right)$} using the standard residual error square, see (\ref{HAM13})} \\ 
\hline
 $1^{st}$   &   0.68778                                             &  0.000066                             &  0.0913                      & 0.2497  \\
 \hline
  $2^{nd}$ &   0.000026                                          &  3.4$\times10^{-6}$               & 0.248721                  & 0.250560 \\
 \hline
 $3^{rd}$   &   2.85$\times10^{-7}$                         &  4.09$\times10^{-10}$           &  0.249868                 & 0.2499944 \\
 \hline
  $4^{th}$  &   8.06$\times10^{-10}$                       &  6.13$\times10^{-11}$           & 0.25000526               & 0.2499996 \\
 \hline
  $5^{th}$  &   1.25$\times10^{-11}$                       &  1.36$\times10^{-14}$           & 0.249999412              &  0.250000027   \\
 \hline
\end{tabular}
\caption{Residual error square for increasing order of the approximation using the least action principle (column 2) or the residual error square method (column 3) with $t=1$ (and $a=\frac{1}{\sqrt{2}}$). The corresponding energies at $t=1$ are reported in columns 4 and 5. $k$ and $\gamma$ are constants of order unity.}
\end{table}

\newpage

\section{Cubic oscillator}
\label{Cubicooscillator}

In this Appendix we compare our HAM method based on the optimization of the action with the standard HAM based on the residual error square minimization for the cubic anharmonic oscillator with nonzero velocity $(\dot{x}_0 \neq 0 )$ and zero velocity $(\dot{x}_0 = 0 )$, respectively.
 
\begin{table}[h!t]
\begin{tabular}{ |p{0.8cm}|p{2.3cm}|p{3.2cm}|p{2.5cm}|p{3.cm}|}
\hline
 & \centering{${{c}_0}$ using LHAM, see (\ref{homotopic3})} &  \centering{$\left| E_{exact} - E\left(t=1, \,\,c_0 \right)\right| $} &${{c}_0}$ using the standard-HAM, see (\ref{HAM13})&$\left| E_{exact} - E\left(t=1, \,\,c_0 \right)\right| $ \\
 \hline
 
  \hline 
 $1^{st}$     & 0.5395      & 0.0184                                                              & -0.9220         & 6.3$\times10^{-4}$                \\
  \hline
 $2^{nd}$   & -0.9668     & 4.0$\times10^{-4}$                                 & -1.0516         & 9.3$\times10^{-6}$              \\
 \hline
$3^{rd}$    & -0.9104     & 8.5$\times10^{-5}$                              & -0.9765          & 2.2$\times10^{-6}$      \\
 \hline
$4^{th}$    & -0.9856     & 1.5$\times10^{-6}$              & -1.0083          & 5.9$\times10^{-8}$     \\
\hline
$5^{th}$    & -0.9917    & 5.4$\times10^{-6}$                & -1.0302          & 6.1$\times10^{-7}$    \\
\hline
\end{tabular}
\caption{The HAM parameter $c_0$ (column 2) with increasing order of the approximation (column 1). In column 3 we report the variation of the energy at time t = 1. Columns 4 and 5 show the HAM parameter $c_0$ and the variation of the
energy using the residual error square method. 
Here, $x_{0} =\frac{1}{\sqrt 2}$ and  $\dot{x}_0=\frac{1}{4}$, i.e, nonzero velocity.}
\label{tab:quarticresd1}
\end{table}

\begin{table}[h!t]
\begin{tabular}{|p{0.8cm}|p{2.3cm}|p{3.0cm}| p{2.8cm}|p{2.8cm}|}
\hline
& {\centering{Residual error square using LHAM, see (\ref{homotopic3}) and (\ref{appendixharm})} }  & { \centering{Standard residual error square, see (\ref{HAM13}) and (\ref{appendixharm})}} & { \centering{$E\left(t=1, \,\,c_0 \right)$ using LHAM, see (\ref{homotopic3})}} & { \centering{$E\left(t=1, \,\,c_0 \right)$ using the standard residual error square, see (\ref{HAM13})}} \\ 
\hline
 $1^{st}$   &   0.1062                                                                      &  0.0002                                  &   -0.113747                                    &   -0.13151 \\
 \hline
  $2^{nd}$ &   0.000015                                                                 &  1.09$\times10^{-6}$               &  -0.1317467                                   &   -0.13215 \\
 \hline
 $3^{rd}$   &   7.2$\times10^{-7}$                                                  &  2.8$\times10^{-9}$                &   -0.1320634                                   &  -0.132146   \\
 \hline
  $4^{th}$  &   3.7$\times10^{-10}$                                                &  1.8$\times10^{-12}$             &    -0.132150450                              &  -0.1321488     \\
 \hline
  $5^{th}$  &   4.6$\times10^{-9}$                                                  &  1.9$\times10^{-10}$              &    -0.132154362                               & -0.1321482 \\
 \hline
\end{tabular}
\caption{Residual error square for increasing order of the approximation using the least action principle (column 2) or the residual error square method (column 3). The corresponding energies at $t=1$ are reported in columns 4 and 5. Here, $x_{0} =\frac{1}{\sqrt 2}$ and  $\dot{x}_0=\frac{1}{4}$, i.e, nonzero velocity.}
\label{tab:quarticresd2}
\end{table}

\begin{table}[h!t]
\begin{tabular}{ |p{0.8cm}|p{2.3cm}|p{3.2cm}|p{2.5cm}|p{3.0cm}|}
\hline
 & \centering{${c_0}$} using LHAM, see (\ref{homotopic3}) &  \centering{$\left| E_{exact} - E\left(t=1, \,\,c_0 \right)\right| $} &${{c}_0}$ using the standard-HAM, see (\ref{HAM13} &$\left| E_{exact} - E\left(t=1, \,\,c_0 \right)\right| $ \\
 \hline
 
  \hline 
 $1^{st}$     &  3.1710          &   0.11508                              & -0.8713             &  2.03$\times10^{-3}$               \\
  \hline
 $2^{nd}$   &  -0.9394          &  9.78$\times10^{-5}$           & -1.0095             &   9.7$\times10^{-4}$                \\
 \hline 
$3^{rd}$    &  -1.0080          & 5.56$\times10^{-5}$             &  -0.9599            &   1.02$\times10^{-5}$              \\
 \hline
$4^{th}$    &  -0.9779          &  2.67$\times10^{-6}$            & -0.9947             &  2.1$\times10^{-7}$                  \\
\hline
$5^{th}$    & -1.0001          & 1.39$\times10^{-7}$              &  -1.0140            &    1.8$\times10^{-8}$                 \\
\hline
\end{tabular}
\caption{
The same as in Table \ref{tab:quarticresd1} but now with $\dot{x}_0=0$, i.e, initial zero velocity.}
\end{table}

\begin{table}[h!t]
\begin{tabular}{|p{0.8cm}|p{2.3cm}|p{3.0cm}| p{2.8cm}|p{2.8cm}|}
\hline
& {\centering{Residual error square using LHAM, see (\ref{homotopic3}) and (\ref{appendixharm})} }  & { \centering{Standard residual error square, see (\ref{HAM13}) and (\ref{appendixharm})}} & { \centering{$E\left(t=1, \,\,c_0 \right)$ using LHAM, see (\ref{homotopic3})}} & { \centering{$E\left(t=1, \,\,c_0 \right)$ using the standard residual error square, see (\ref{HAM13})}} \\ 
\hline
 $1^{st}$   &   0.40540                                     &  7.1$\times10^{-4}$               &  -0.11508                        & -0.09886   \\
 \hline
  $2^{nd}$ & 1.7$\times10^{-5}$                     & 2.70$\times10^{-8}$              &  -0.09992                        & -0.100893  \\
 \hline
 $3^{rd}$   & 8.4$\times10^{-8}$                    & 1.1$\times10^{-8}$                & -0.1009544                      & -0.100888  \\
 \hline
  $4^{th}$  & 2.2  $\times10^{-10}$                & 2.6$\times10^{-12}$               & -0.10090154                    & -0.1008990 \\
 \hline
  $5^{th}$  & 9.94$\times10^{-13}$                &  1.70$\times10^{-13}$           &  -0.10089872                    &-0.1008988  \\
 \hline
\end{tabular}
\caption{
The same as in Table \ref{tab:quarticresd2} but now with $\dot{x}_0=0$, i.e, initial zero velocity.}
\end{table}



\newpage

\bigskip


\end{document}